\documentclass[pre,aps,showpacs,twocolumn,supergroupedaddress,epsfig,amsmath,amssymb,eqsecnum]{revtex4}
\usepackage{epsfig,amsmath,amssymb,bm,epsf,graphics,psfrag,verbatim,subfigure,framed}
\usepackage[usenames,dvipsnames]{color}
\usepackage{bbm}
\usepackage{mathrsfs}
\usepackage{amsfonts}
\usepackage{color}
\usepackage{pbox}
\def\newblock{\hskip .11em plus .33em minus .07em}

\def\erf{\text{erf}}
\def\tf{\widetilde{f}}
\def\tv{\widetilde{v}}
\def\tp{\widetilde{p}}
\def\width{b}
\def\xvp{\mathbf{x}_\perp}

\def\np{\nabla_\perp}

\def\gouy{\mu}

\def\stiff{\mathfrak{K}}

\def\stiff{\kappa}
\def\xp{x_\perp}

\def\np{\nabla_\perp}
\def\tV{\widetilde{V}}
\def\tH{\widetilde{H}}
\def\tw{\widetilde{w}}

\def\wp{\widetilde{p}}
\def\hard{\lambda}
\def\bjerrum{\ell_{{\mathrm B}}}
\def\boltzmann{k_{{\mathrm B}}}
\def\tl{\widetilde{\ell}_0}
\def\tls{\widetilde{\ell}^\ast}
\def\tS{\widetilde{S}}

\newcommand{\be}{\begin{equation}}
\newcommand{\ee}{\end{equation}}
\newcommand{\ba}{\begin{eqnarray}}
\newcommand{\ea}{\end{eqnarray}}
\newcommand{\bw}{\begin{widetext}}
\newcommand{\ew}{\end{widetext}}

\newcommand{\Qv}{{\bm{Q}}}

\newcommand{\qv}{{\bm{q}}}

\newcommand{\bing}[1] {{\color{black} #1 }}

\begin{document}

%\title{Interactions of mutually non-penetrating fluid membranes}
%\title{Effective interactions between steric fluid membranes}
\title{Effective interactions between fluid membranes}
\author{Bing-Sui Lu}
\email{bing-sui.lu@fmf.uni-lj.si}
\author{Rudolf Podgornik}
\email{rudolf.podgornik@fmf.uni-lj.si}
\affiliation{Department of Physics, Faculty of Mathematics and Physics, University of Ljubljana, Jadranska ulica 19, SI-1000 Ljubljana, Slovenia \\ and Department of Theoretical Physics, J. Stefan Institute, 1000 Ljubljana, Slovenia.
}
%\author{A. C. Maggs}
%\affiliation{Physico-chimie th\'{e}orique--Gulliver, ESPCI-CNRS, 10 rue Vauquelin 75005 Paris, France}

\date{\today}

\begin{abstract}
A self-consistent theory is proposed for the general problem of interacting undulating fluid membranes subject to the constraint that they do not interpenetrate. \bing{We implement the steric constraint via an exact functional integral representation, and through the use of a saddle-point approximation transform it into a novel effective steric potential.} The steric potential is found to consist of two contributions: one generated by zero mode fluctuations of the membranes, and the other by thermal bending fluctuations. For membranes of cross-sectional area $S$, we find that the bending fluctuation part scales with the inter-membrane separation $d$ as $d^{-2}$ for $d \ll \sqrt{S}$, but crosses over to $d^{-4}$ scaling for $d \gg \sqrt{S}$, whereas the zero mode part of the steric potential always scales as $d^{-2}$. For membranes interacting exclusively via the steric potential, we obtain \bing{closed-form} expressions for the effective interaction potential and for the rms undulation amplitude $\sigma$, which becomes small at low temperatures $T$ and/or large bending stiffnesses $\kappa$. Moreover, $\sigma$ scales as $d$ for $d \ll \sqrt{S}$, but saturates at $\sqrt{k_{{\rm B}} T S/\kappa}$ for $d \gg \sqrt{S}$. In addition, using variational Gaussian theory, we apply our self-consistent treatment to study inter-membrane interactions subject to three different types of potential: (i)~the Moreira-Netz potential for a pair of strongly charged membranes with an intervening solution of multivalent counterions, (ii)~an attractive square well, (iii)~the Morse potential, and (iv)~a combination of hydration and van der Waals interactions. 
\end{abstract}

\maketitle

\section{Introduction}
\label{sec:introduction}

Lipid membranes are two-dimensional fluids lacking in-plane shear elasticity, and the only elastic penalty they experience comes from changes in curvature~\cite{nagle_review}. Thus, they undergo vigorous thermally activated undulations, and a pair of such membranes brought into proximity will experience an osmotic pressure generated by the reduction of phase space available for the membranes to fluctuate. The problem of determining this osmotic pressure is a classic one, going back to Helfrich~\cite{helfrich}, and over the past four decades different methods have been proposed to treat it, employing various degrees of heuristic argument, field theoretic tools, and functional renormalization group (FRG) techniques (see, e.g., Refs.~\cite{helfrich_servuss, janke_kleinert, evans_parsegian,podgornik_parsegian, manghi, david, farago, freund, mecke, sornette_ostrowsky, sornette}). A major difficulty that has dogged research in this area is associated with the proper treatment of the steric constraint, i.e., the constraint that a pair of membranes may not penetrate each other. One response was to replace the problem of implementing the steric constraint at the level of the partition function by an {\sl Ansatz} that the root mean square (rms) fluctuation amplitude (or the amplitude of ``roughness") $\sigma$ of a membrane is of the order of the inter-membrane separation $d$, viz., $\sigma^2 = \hat{\mu} d^2$, where $\hat{\mu}$ is some numerical constant. One may interpret the mean square fluctuation as the inverse curvature of a harmonic potential well. By integrating out the bending fluctuations and applying $\sigma^2 = \hat{\mu} d^2$, one then obtains a steric potential of the form $V_H = c_{{\rm fl}}(\boltzmann T)^2/\stiff d^2$, which is known as the Helfrich fluctuation potential~\cite{helfrich_servuss, janke_kleinert}, with $\stiff$ the curvature stiffness of the membrane. From this potential one derives the fluctuation osmotic pressure, one that is entirely generated by constrained thermal fluctuations of the membranes. It has been argued (see, e.g., \cite{helfrich,helfrich_servuss, david}) in the past that $c_{{\rm fl}}$ is a universal number~\cite{footnote_frg}. On the other hand, one may reasonably expect that the rms fluctuation $\sigma$, being the effect of thermally activated undulations of the membrane, vanishes as $T\rightarrow 0$ and/or $\stiff\rightarrow \infty$, and accordingly $\hat{\mu}$ should depend on $T/\stiff$. Moreover, as $d$ increases, the behavior of $\sigma$ should cross over to that of a freely undulating membrane, i.e., scale as $\sqrt{S}$ (where $S$ is the transverse projected area of the membrane; cf. Ref.~\cite{helfrich_servuss}), and thus $\sigma$ should also be a (possibly nonlinear) function of $d/\sqrt{S}$.

In the foregoing paragraph we have described the problem of the steric potential generated purely by the thermal undulations of a pair of membranes and the hard-wall constraint. In more realistic systems (including those of biological interest), the steric potential is modified by the presence of other interactions, both short- and long-range, and these may include van der Waals, electrostatic (if the membranes are charged), charge regulation and/or hydration forces. Problems of relevant concern include: (i)~what is the behavior or form of the effective interaction potential? (ii)~How does the equilibrium inter-membrane separation change as a function of external osmotic pressure? (iii)~In the case where the interaction is attractive at long range but repulsive at short range, is there an unbinding transition at zero external osmotic pressure? If so, what is the order of the unbinding transition and the threshold value of the (attractive) interaction strength for the membranes to unbind? Besides the presence of additional interactions in real systems, the membranes themselves can be multi-component, e.g., consist of a mixture of lipids and lateral inclusions that may undergo phase separation~\cite{komura_andelman}, and the solution may also contain polymers (such as proteins) which then interact with the membranes and anchor or adsorb onto the membrane surfaces~\cite{benhamou}. 

Theoretical treatments of such problems have traditionally fallen into four classes: mean-field or Flory-Huggins-type approaches (see, e.g., \cite{wennerstrom,milner_roux,helfrich1,komura_andelman}), variational Gaussian approximation (VGA) schemes (see, e.g., \cite{evans_parsegian, podgornik_parsegian, benhamou,manghi}), FRG-based methods (see, e.g., \cite{lipowsky1,lipowsky2, lipowsky3,lipowsky4,lipowsky5,lipowsky6, lipowsky1990,david_leibler, sornette}), and Monte Carlo simulations of different flavours (see, e.g., \cite{lipowsky_zielinska, netz_lipowsky1,netz_lipowsky, netz_unbinding, gouliaev1, gouliaev2}). For the simpler case of a pair of single-component membranes interacting via a direct potential that consists of a long-range attractive van der Waals tail and a short-range repulsive hydration potential, mean-field theory (MFT) based on the simple addition of the direct potential with the fluctuation-induced steric potential predicts a first order unbinding transition at a critical strength of the Hamaker coefficient~\cite{wennerstrom}, whereas a Flory-Huggins-type theory, modeled after the van der Waals theory of liquid-gas condensation, predicts a continuous unbinding transition at a critical strength of the Hamaker coefficient $W=W_c$ and scaling behavior $d \sim |W-W_c|^{-1}$~\cite{milner_roux}. Such an unbinding transition in three spatial dimensions has also been predicted by FRG-based calculations~\cite{lipowsky1, lipowsky2}. 
%which rely on an effective potential computed at harmonic order in fluctuations of $d$, subject to a hard-wall boundary condition at $d=0$, and an assumption that the fluctuation-induced steric interaction decays as $d^{-2}$ for large $d$. 
On the other hand, experimental reports that the unbinding transition in lipid bilayer systems is of first order~(see e.g. Refs.~\cite{vogel, fontell, khan,larsson}) have been more than those that report the unbinding transition to be of second order~(see e.g. Ref.~\cite{mutz}). 

Using a novel approach based on a self-consistent treatment of the hard-wall constraint, we revisit the problems described in the foregoing paragraphs. The requirement of self-consistency is realized by implementing the hard-wall constraint at the level of the partition function via the use of Heaviside function. By making use of a representation first proposed by Panyukov and Rabin~\cite{panyukov_rabin} in a rather different context, the steric constraint can be transformed into terms of an effective interaction Hamiltonian. By considering small undulations of the membrane, we \emph{derive} \bing{closed-form} self-consistent expressions for the mean square fluctuation $\sigma^2$ and the equilibrium inter-membrane separation distance, which precisely describe how the strength of fluctuations is related to both temperature $T$ and the curvature stiffness $\stiff$ of the fluctuating membrane \cite{nagle_review}, as well as the average inter-membrane separation distance $d$. This effective Hamiltonian also enables one to see the precise mechanism in which the fluctuation osmotic pressure emerges from fluctuation and steric forces, and shows a crossover of scaling behavior from $\sigma^2 \sim \sqrt{\boltzmann T/\stiff} d^2$, when a pair of membranes are close to one another, to $\sigma^2 \sim (\boltzmann T/\stiff) S$ (where $S$ denotes the transverse projected area of each membrane), when the membranes are far apart.
%~\cite{footnote:crumpling}.  

Our self-consistency requirement is reinforced for the case of real systems (i.e., where the direct potential now includes non-Gaussian terms stemming from electrostatic, van der Waals, charge regulation, and/or hydration interactions) by our use of the VGA. The VGA captures the fact that the magnitude of the fluctuation-induced steric potential is influenced by the direct potential, and vice versa \cite{evans_parsegian, podgornik_parsegian, petrache}. Mathematically this translates into the problem of solving a coupled pair of (typically nonlinear) equations for the rms fluctuation amplitude and the average separation (or more generally speaking the fluctuation correlator of an observable and the average value of that observable). It is known that the VGA becomes exact as the co-dimension of the manifold (which is equal to the number of components of the vector field representing the transverse fluctuations of the manifold) tends to infinity~\cite{mezard_parisi}. 

The VGA can be regarded as being complementary to FRG techniques.  The FRG is essentially an asymptotic method which has been used to study problems relevant to critical behavior and the associated scaling exponents (i.e., the ``shape" of the effective interaction), addressing the limiting behavior of the interaction potential in the regions of small and large inter-membrane separations. 
%The FRG is essentially an asymptotic method which can be of use in studying problems to do with critical behavior and the associated scaling exponents (i.e., the ``shape" of the effective interaction) in the limits of small and large inter-membrane separations. 
%predicts the scaling behavior (i.e., the ``shape") of the effective interaction potential in the near- and far-field limits, and comes in most handy as a tool for predicting the order of a phase transition and the associated scaling exponents. 
%and thus cannot elucidate how the key physical parameters (such as temperature, membrane stiffness, Hamaker strength, and lipid bilayer thickness) influence the behavior of the effective interaction potential, 
%As such it is not very useful for systems that are near or far away from criticality, nor does it elucidate the physical mechanism of effective interactions (i.e., how these interactions emerge from more microscopic ones).
As such it is less useful as a tool for making predictions about the system's behavior at \emph{intermediate} separations, which are the typical length scales of biologically relevant systems. At such length scales there is usually a number of competing interactions of comparable strength, and the use of asymptotic methods is not of much practical value~\cite{evans_ninham}. One of the primary interests in membrane biophysics is also to elucidate how the effective mesoscale inter-membrane interaction, accessible via detailed osmotic stress experiments on multilamellar lipid systems \cite{rand_parsegian, petrache, leneveu, nagle_review0, pabst}, arises from details of more microscopic forces, an aspect that is not captured by the FRG-based methods at all. These aspects can be adequately addressed by a VGA-based approach. 
%would enable one to predict not only the separation dependence, but also the strength of the effective interaction once the physical parameters are known. 
Moreover, the VGA-based approach is comparatively simple to implement and the physical mechanism shines through the formalism more transparently. 

Among our main motivations here is also to analyze the behavior of the electrostatic correlation interaction in the strong coupling limit \cite{Perspective} that has been already observed between stiff charged silica surfaces \cite{Sivan} and is now being investigated also in the case of soft, fluctuating membranes \cite{pabst}. It seems to us important to be able to understand the difference between the experimentally measured ion correlation effect between stiff and soft interfaces and to isolate in what way the membrane fluctuations change the expected separation behavior of the strong coupling interactions. Our analysis is the first in this direction and should be helpful to understand the experimental data when available.

Our Paper is divided into the following sections. Section II introduces the problem of determining the osmotic  pressure of a freely undulating membrane near a hard wall, and describes how the hard-wall constraint can be transformed into effective potential terms via the Panyukov-Rabin representation. The system considered is a pair of membranes subject only to hard-wall repulsion and undergoing thermal bending fluctuations. The first main result of this section is Eq.~(\ref{eq:sigma_exact}) for the rms fluctuation amplitude of a  membrane near a hard wall, which interpolates between the scaling behavior of a confined membrane at short separations and that of a freely undulating membrane at large separations. The second main result is Eq.~(\ref{eq:helf-like}), which expresses the steric potential as the sum of contributions from thermal bending and zero mode fluctuations.  In Sec.~III, we introduce a variational method due to Feynman and Kleinert~\cite{feynman_kleinert}, which we apply to study the behavior of membranes in more realistic systems, i.e., ones that are characterized by effective interaction potentials of non-Gaussian form (specifically, an attractive square well potential, the Morse potential, the Moreira-Netz potential, and a combination of van der Waals and hydration energies). In Sec IV we present our discussion and conclusions.

\section{Membrane near a hard wall}
\label{sec:one_hard_wall}

The fluctuation osmotic pressure of a pair of mutually impenetrable membranes of bending stiffnesses $\stiff_1$ and $\stiff_2$ \cite{nagle_review} is equivalent to that of a membrane of stiffness $\stiff \equiv \stiff_1 \stiff_2/(\stiff_1+\stiff_2)$ near a hard (i.e., impenetrable) wall. We shall thus consider a single membrane fluctuating near a hard wall as our prototypical system. To maintain the membrane at constant average distance from the wall, an external osmotic  pressure $P$ that is equal and opposite to its fluctuation osmotic pressure has to be applied. We fix the position of the wall at the origin of the $z$-axis, and a point on the membrane shall have transverse coordinates $\xvp = (x,y)$ and occupy a position $\ell(\xvp)$ on the $z$-axis. 
%Our prototypical system consists of a hard wall and a fluctuating membrane, with an external osmotic pressure $P$ applied on the membrane. We fix the origin of the $z$-axis at the position of the wall, and a point on the membrane with transverse coordinates $\xvp = (x,y)$ occupies a position $\ell(\xvp)$ along the $z$-axis. 
The space between the wall and the membrane could be empty or filled with electrolyte solution. \bing{The membrane's total energy is given by}
\be
\tH = \int \!d^2 \xp \big( \frac{\stiff}{2} (\np^2 \ell(\xvp))^2 + \tV(\ell(\xvp)) \big),
\label{eq:interaction}
\ee
\bing{where the interaction potential is given by}
\be
\tV(\ell(\xvp)) = \tw(\ell(\xvp)) + P \, \ell(\xvp).
\label{eq:tV}
\ee
Denoting the (geometric) mean separation by $\ell_0$, defined by $\int \! d^2\xp \, \ell(\xvp) = \ell_0$, we can write
\be
\ell(\xvp) = \ell_0 + \delta\ell(\xvp) = \ell_0 + \int \!\! \frac{d^2 Q}{(2\pi)^2} \, e^{i\Qv\cdot\xvp}\delta\ell_Q, 
\ee 
where the wave-vector integral excludes the $\Qv = {0}$ mode, and $\delta\ell_Q$ are the Fourier modes of the deviation $\delta\ell(\xvp)$.  Very soon [cf. Eq.~(\ref{eq:H_eff})] we shall see the beauty of such a decomposition that lies at the heart of the Feynman-Kleinert variational theory: the terms that are linear in fluctuation vanish from the Hamiltonian (owing to the integral over the transverse projected area of the membrane), and the resulting integration over the fluctuating fields is trivially Gaussian.  

There are three terms in Eqs.~(\ref{eq:interaction}) and (\ref{eq:tV}). The first is the free energy associated with bending undulation modes of the membrane.  The second term $\tw$ represents the direct potential between the wall and the membrane, \bing{arising} from forces of electrostatic and non-electrostatic (such as hydration) origin. In general the direct potential has a non-trivial form and is non-Gaussian in fluctuations. 
%Owing to such structural non-triviality, the fluctuation osmotic pressure cannot be easily calculated for membranes that experience electrostatic interactions. 
The third term represents the effect of applying an external osmotic  pressure $P$ on the fluctuating membrane. Experimentally, for membranes in solution the external osmotic pressure can be controlled by using the classic method of Rand and Parsegian~\cite{rand_parsegian}, which is to vary the concentration of polymer (such as dextran~\cite{leneveu}, polyvinylpyrrolidone~\cite{nagle_review0} or polyethylene glycol~\cite{pabst}) added to extract water from between the membranes. 
%The sign is positive, because on increasing the osmotic pressure, the system should respond by decreasing the separation $\ell$. 
Theoretically, the osmotic pressure term can also be interpreted as a source field that one uses to differentiate the logarithm of $Z$ in order to obtain the equilibrium separation distance $\langle \ell_0 \rangle$. 

\subsection{Hard-wall constraint: functional representation}
 \begin{figure}
		\includegraphics[width=0.41\textwidth]{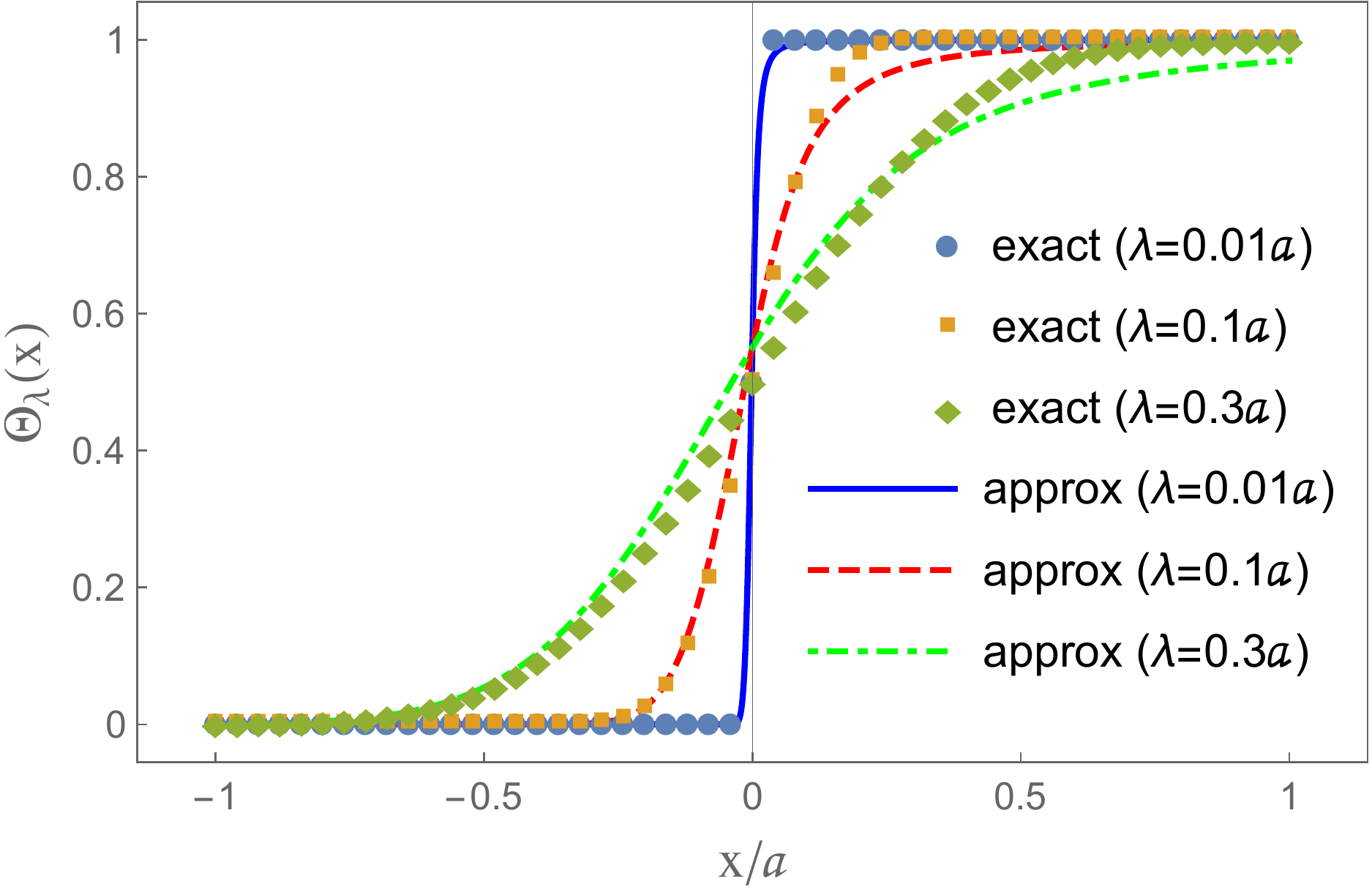}
	\caption{Comparison of plots of $\Theta_\hard(x)$ generated using the exact function [Eq.~(\ref{eq:Theta_exa}); behavior is described by the blue-grey circles, yellow squares, and green diamonds for $\lambda=0.01a$, $0.1a$ and $0.3a$ respectively, where $a$ is a microscopic lengthscale] and the saddle point approximation [Eq.~(\ref{eq:sp00}); behavior is displayed by the blue, red dashed, and green dot-dashed lines for $\lambda=0.01a$, $0.1a$ and $0.3a$ respectively]. The agreement is quite good, with the accuracy improving for smaller values of $\lambda$.}
\label{fig:heaviside}
\end{figure}

The partition function of a fluid membrane near a hard wall is then given by 
\be
Z = \prod_{\{ \xvp \}}\int_{-\infty}^{\infty} d\ell(\xvp) \, \Theta(\ell(\xvp)) \, e^{-\beta \tH}
\label{eq:Z_unconstrained0}
\ee
where $\beta=1/\boltzmann T$ and $\Theta(\ell(\xvp))$ is the Heaviside function which is equal to unity if $\ell(\xvp)>0$ and zero if $\ell(\xvp)<0$. Taking up an idea in Ref.~\cite{panyukov_rabin}, we express the Heaviside function in terms of \bing{the Dirac delta function:}   
\be
\Theta(\ell(\xvp)) = \int_0^\infty d\Lambda \, \delta(\Lambda - \ell(\xvp)).
\label{eq:heaviside0}
\ee 
Next, we use a Gaussian representation for the delta function, and write the Heaviside function as 
\ba
\Theta(\ell(\xvp)) &=& \lim_{\hard \rightarrow 0} \Theta_\hard(\ell(\xvp)) 
\nonumber\\
&\equiv& \lim_{\hard \rightarrow 0}  
\int_0^\infty \!\!\!\! d\Lambda \frac{1}{\sqrt{2\pi \hard^2}} e^{-\frac{(\Lambda-\ell(\xvp))^2}{2\hard^2}}.
\label{eq:heaviside_mu}
\ea
The limit $\hard \rightarrow 0$ is a mathematical idealization for an ``infinitely hard" wall. However, for real membranes, the boundary region is not infinitely sharp, but has a certain extent which is set by the lengthscale of the lipid \bing{headgroup} size $a$. The parameter $\hard$ reflects the size of this boundary region. We can thus write 
\be
\hard = c \, a,
\ee
where $c$ depends on the microscopic details of the chemical make-up of the membrane. 

By making a change of variables $\Lambda = \frac{1}{2}u^2$ and promoting a prefactor $u$ into the exponent by means of a logarithm, Eq.~(\ref{eq:heaviside_mu}) can be expressed in the form
\be
\Theta(\ell(\xvp)) = \lim_{\hard \rightarrow 0} \int_{-\infty}^\infty du \, \frac{1}{\sqrt{2\pi \hard^2}} e^{f_\hard(u, \ell(\xvp))}
\label{eq:heaviside1}
\ee 
where 
\be
f_\hard(u, \ell) \equiv \frac{1}{2}\ln u^2 -\frac{(\frac{1}{2}u^2-\ell)^2}{2\hard^2}.
\label{eq:Theta_exa}
\ee
\bing{We have obtained an {\em exact} functional integral representation of the steric constraint for each point of the membrane.}  
In contrast to $\Lambda$, the variable $u$ now runs over the unbounded interval $(-\infty, \infty)$. 
%This eliminates the complication arising from functional integrals over fluctuations of $\Lambda(\xvp)$ that have non-zero and finite lower bounds [**Add a footnote here explaining this technical issue]. 
Using Eq.~(\ref{eq:heaviside1}), we can express the partition function in Eq.~(\ref{eq:Z_unconstrained0}) as 
\ba
Z &=&  \lim_{\hard\rightarrow 0}  
\prod_{\{ \xvp \}}\int_{-\infty}^{\infty} d\ell(\xvp) 
\int_0^\infty \frac{d\Lambda(\xvp)}{\sqrt{2\pi \hard^2}} 
\nonumber\\
&&\quad e^{-\int\frac{d^2\xp}{a^2} \frac{(\Lambda(\xvp) - \ell(\xvp))^2}{2\hard^2}} \, e^{-\beta \tH}
\nonumber\\
&=&  \lim_{\hard\rightarrow 0}  
\prod_{\{ \xvp \}}\int_{-\infty}^{\infty} d\ell(\xvp) 
\int_{-\infty}^\infty \frac{du(\xvp)}{\sqrt{2\pi \hard^2}} 
\nonumber\\
&&\quad 
e^{\int\frac{d^2\xp}{a^2} f_\hard(u(\xvp), \ell(\xvp))} \, e^{-\beta \tH}
\label{eq:Z_g}
\ea
In the functional integral expression for $Z$, we have introduced a microscopic cut-off length scale $a$, in order to convert the sum over a two-dimensional lattice of points into an integral. The length scale $a$ is set by the size of the membrane's molecular constituents, e.g., the diameter of the lipid headgroup. 

\subsection{Steric potential: saddle-point approximation}
Let us now make the saddle point approximation to $e^{f_\hard}$. It is given by 
\be
e^{f_\hard(u, \ell)} \approx \sqrt{\frac{2\pi}{|f_\hard''(u_{{\rm sp}}, \ell)|}} e^{f_\hard(u_{{\rm sp}}, \ell)} 
\label{eq:sp_appr}
\ee
where $u_{{\rm sp}}$ solves the saddle point equation
\ba
&&f_\hard'(u_{{\rm sp}}, \ell)=0
\nonumber\\
&&\Rightarrow u_{{\rm sp}}^2(\xvp) = \ell(\xvp) + \sqrt{\ell(\xvp)^2 + 2\hard^2} 
\ea
We find that the saddle point approximation to $f_\hard(u)$ is given by $f_\hard^\ast$, where 
\ba
f_\hard^\ast &=& 
\frac{1}{2}\ln(2\pi) -\frac{1}{2}\ln\left\{ \frac{2(2\hard^2+\ell(\ell+\sqrt{2\hard^2+\ell^2}))}{\hard^2(\ell+\sqrt{2\hard^2+\ell^2})} \right\} 
\nonumber\\
&&+\frac{1}{2}\ln(\ell+\sqrt{2\hard^2+\ell^2}) - \frac{1}{8\hard^2} (\ell-\sqrt{2\hard^2+\ell^2})^2,    
\nonumber\\
\label{eq:sp00}
\ea
where we have written $\ell \equiv \ell(\xvp)$ to lighten our notation. 
In Fig.~\ref{fig:heaviside}, we compare the saddle point approximation of $\Theta_\hard$ (viz., $e^{f_\hard^\ast}$) with the exact function [Eq.~(\ref{eq:Theta_exa})]. The agreement is excellent, especially for smaller values of $\lambda$. We can thus approximate $Z$ in Eq.~(\ref{eq:Z_g}) by its saddle-point value, viz., 
\be
Z^\ast = \lim_{\hard\rightarrow 0}  
\prod_{\{ \xvp \}}\int_{-\infty}^{\infty} d\ell(\xvp)  
e^{\int\frac{d^2\xp}{a^2} f_\hard^\ast(\ell(\xvp))} \, e^{-\beta \tH}
\label{eq:Z_g^*}
\ee
On going from Eq.~(\ref{eq:Z_unconstrained0}) to Eq.~(\ref{eq:Z_g^*}), we have effectively moved from a partition function involving $\ell(\xvp)$ subject to a hard-wall constraint to a partition function where $\ell(\xvp)$ is \emph{unconstrained}, but the steric condition is enforced \emph{energetically} through terms in an effective Hamiltonian. In our calculational steps we have not made any assumptions and made use of only one approximation, viz., the saddle-point approximation of Eq.~(\ref{eq:sp_appr}). 

As $\hard$ is small, we can expand the above to leading order in $\hard$:
\ba
f_\hard^\ast &\!=\!& \frac{-\ell^2+\ell\sqrt{\ell^2}}{4\hard^2}  
-\frac{1}{4} + \frac{\sqrt{\ell^2}}{4\ell} 
+ \frac{1}{2}\ln\left[ \frac{\ell+\sqrt{\ell^2}}{2\ell}\right] 
\nonumber\\
&&
-\frac{(\ell+5\sqrt{\ell^2})\hard^2 }{8\ell\sqrt{\ell^2}(\ell+\sqrt{\ell^2})}
+ \ln \hard+ O(\hard^3)
\label{eq:f_small}
\ea
where we have neglected a constant term $\frac{1}{2}\ln(2\pi)$. 
We see that terms for which $\ell$ is negative diverge to negative infinity, and so $e^{f_\hard^\ast}$ goes to zero, which reflects the zero probability of finding the membrane inside the wall. 

We now focus on the case of membranes undergoing thermally excited undulations with \emph{small} amplitude. Let us write $\ell(\xvp) = \ell_0 + \delta \ell(\xvp)$, where $\ell_0$ is the \emph{geometric} (not thermal) mean defined by $\ell_0 \equiv (1/S)\int d^2\xp \ell(\xvp)$, and $\delta\ell(\xvp)$ are small deviations around $\ell_0$. In Fourier space, $\ell_0$ is the zero wave-vector mode and  
\be
\delta\ell(\xvp) = {\sum_{\Qv}}^\prime e^{i\Qv\cdot\xvp} \delta\ell_\Qv
\ee
where $\Qv$ is a two-dimensional wave-vector conjugate to $\xvp$, and the prime denotes the exclusion of the zero wave-vector mode from the wave-vector sum. 
Both $\ell_0$ and $\{ \delta\ell_\Qv \}$ [or $\delta\ell(\xvp)$] are independent thermally fluctuating variables that take values from the interval $(-\infty,\infty)$ [cf. Eq.~(\ref{eq:Z_g})].  
By performing an expansion to quadratic order in $\delta\ell(\xvp)$, Eq.~(\ref{eq:f_small}) becomes 
\ba
f_\hard^\ast &\approx& -\frac{(\ell_0+5\sqrt{\ell_0^2})\hard^2 }{8\ell_0\sqrt{\ell_0^2}(\ell_0+\sqrt{\ell_0^2})}
+\frac{(\ell_0+5\sqrt{\ell_0^2})\hard^2}{4(\ell_0^2)^{3/2}(\ell_0+\sqrt{\ell_0^2})} \delta\ell
\nonumber\\
&&- \frac{(2\ell_0+\sqrt{\ell_0^2})(\ell_0+5\sqrt{\ell_0^2})\hard^2}{4\ell_0^3\sqrt{\ell_0^2}(\ell_0+\sqrt{\ell_0^2})^2}\delta\ell^2. 
\label{eq:f_hard}
\ea
For negative values of $\ell_0$ the leading-order term of the above expression diverges to negative infinity, which means that the geometric average position of the membrane cannot move inside the wall. 
%The small $\delta\ell$ approximation has thus replaced the ``exact" impenetrability of the membrane [as described by Eq.~(\ref{eq:f_small})] by an ``average" impenetrability.  
Under this approximation, we can effectively restrict the range of values that $\ell_0$ takes in the partition function to the positive interval $(0,\infty)$, and re-express Eq.~(\ref{eq:f_hard}) as
\be
f_\hard^\ast \approx - \frac{3\hard^2}{8\ell_0^2} + \frac{3\hard^2}{4\ell_0^3}\delta\ell(\xvp) 
-\frac{9 \hard^2}{8 \ell_0^4} (\delta\ell(\xvp))^2. 
\ee
We note that the term linear in $\delta\ell(\xvp)$ will vanish when we integrate over the transverse projected area. The saddle point approximation to $Z$ can thus be written as 
\be
Z \approx 
\int_{0}^{\infty} \!\!\! d\ell_0 \! \prod_{\{ \xvp \}} \! \int_{-\infty}^{\infty} \!\!\! d\delta\ell(\xvp) 
e^{-\beta \! \int \! d^2 \xp ( \frac{\stiff}{2} (\np^2 \ell(\xvp))^2 + V )},
\label{eq:Z_eff}
\ee
where $V$ is now an effective potential defined by  
\be
V = w(\ell(\xvp)) + P\ell(\xvp) 
\label{eq:V_eff}
\ee
and
\be
w(\ell(\xvp))= \tw(\ell(\xvp)) +
\frac{9 \boltzmann Tc^2}{8 \ell_0^4} (\delta\ell(\xvp))^2 + 
 \frac{3 \boltzmann T c^2}{8 \ell_0^2}.
\label{eq:H_eff}
\ee
%where $c \equiv \lambda/a$. 
In the above, we have removed a prefactor $(2\pi\hard^2)^{-S/2a^2}$. This can be done, as the prefactor simply corresponds to a shift of the effective free energy by an (infinite) constant. 

The prefactor of the term quadratic in $\delta\ell$ can be interpreted as an effective compression modulus generated by the steric force acting between the membrane and the wall. The interaction potential $w$ thus consists of two contributions: a \emph{soft} potential $\tw$ that describes the longer-range (and coarse-grained) interactions of the membranes, and a \emph{hard} potential which arises from the steric constraint. Such a decomposition into soft and hard contributions is possible because we are working at an already coarse-grained/mesocopic level; otherwise all contributions (including steric effects) are reducible to electrostatic ones. By rewriting the steric constraint in terms of delta functions [Eqs.~(\ref{eq:heaviside0}) and \ref{eq:heaviside1})] and applying a saddle point approximation on $u$, we were able to transform the partition function for a \emph{sterically constrained} membrane into one for an \emph{unconstrained} membrane, with the effects of the steric constraint accounted for by terms in an \emph{effective} interaction potential $w$ [cf. Eq.~(\ref{eq:H_eff})]. This also completes the program set out in Ref. \cite{podgornik_parsegian} but not carried out to its full implementation.
%\bing{The sterically generated contributions in Eq.~(\ref{eq:H_eff}) are to be integrated over the transverse projected surface area of the membrane in the effective Hamiltonian. These contributions are derived from a Gaussian approximation (i.e., correct to quadratic order in $\delta\ell$) to $f_\hard(u_{{\rm sp}}, \ell)$ in Eq.~(\ref{eq:sp00}), and our results for the mean square fluctuation $\sigma^2$ should be strictly limited to the regime where $S\sigma^2/\ell_0^4 \lesssim 1$ or $\sigma^2 \lesssim \ell_0^4/S$.} 

Two observations can immediately be made: (i)~the energy diverges as the mean separation $\ell_0$ tends to zero, which reflects the fact that the membrane is unable to penetrate the wall, and (ii)~the prefactor of the fluctuation term also diverges as $\ell_0 \rightarrow 0$, which reflects the fact that the fluctuations of the membrane must also be suppressed as it reaches the wall.  

\subsection{Application: steric and fluctuation forces}
     
Let us consider a pair of membranes that interact via only steric and fluctuation forces. As we have already noted, the two-membrane system can be recast as a membrane interacting with a hard wall. This problem admits of a \bing{closed-form} solution that describes how $\sigma^2$ depends on $\ell_0$. We set the soft potential to zero: $\tw =  0$. The hard-wall and fluctuation-induced repulsion from the wall has to be balanced by an external osmotic pressure if the membrane is to remain at a finite distance from the wall. For the potential including only the sterically generated interactions, Eq.~(\ref{eq:H_eff}) assumes the form
\be
w(\ell(\xvp))= \frac{9 \boltzmann Tc^2}{8 \ell_0^4} (\delta\ell(\xvp))^2 + 
 \frac{3 \boltzmann T c^2}{8 \ell_0^2}.
\ee
The functional integration over $\delta\ell$ in $Z$ is thus Gaussian and can readily be performed. Using Eq.~(\ref{eq:Z_eff}) and the definition $\sigma^2 \equiv \langle (\delta\ell(\xvp))^2 \rangle_{\delta\ell}$ (where $\langle \ldots \rangle_{\delta\ell}$ denotes averaging over $\delta\ell$ using $Z$ for a given $\ell_0$) yields for the mean square fluctuation amplitude [cf. also Eq.~(\ref{eq:ms_fluct_exact})]
\be
\sigma^2 = \frac{\ell_0^2}{\bing{12}c}\sqrt{\frac{\boltzmann T}{\stiff}}\left( 1 - \frac{2}{\pi}\tan^{-1}\left( \frac{2\ell_0^2}{3cS}\sqrt{\frac{\stiff}{\boltzmann T}} \right) \right).
\label{eq:sigma_exact}
\ee
Comparing with the Ansatz $\sigma^2 = \hat{\mu} d^2$ of Ref.~\cite{helfrich} (where $d=\ell_0$), we see that $\hat{\mu}$ has the following structure: 
\be
\hat{\mu} = \frac{1}{\bing{12}c}\sqrt{\frac{\boltzmann T}{\stiff}}\left( 1 - \frac{2}{\pi}\tan^{-1}\left( \frac{2\ell_0^2}{3cS}\sqrt{\frac{\stiff}{\boltzmann T}} \right) \right).
\ee
In accordance with the expectations described in \bing{the} Introduction, our calculation has revealed the structure behind the prefactor $\hat{\mu}$, showing it to be a nonlinear function of $\boltzmann T/\stiff$ and $\ell_0/\sqrt{S}$. To understand the properties of $\hat{\mu}$, we look at two limiting cases: the regimes of small and large inter-membrane separations. 
Defining the dimensionless variables $\tl \equiv \ell_0/d^\ast$ and $\widetilde{\sigma} \equiv \sigma/d^\ast$, where $d^\ast \equiv \sqrt{3c}(\boltzmann T)^{1/4}\sqrt{S}/(4\stiff)^{1/4}$, the above equation can be put in dimensionless form: 
\be
\widetilde{\sigma}^2 = \frac{1}{\bing{12}c}\sqrt{\frac{\boltzmann T}{\stiff}}\left( 1-\frac{2}{\pi}\tan^{-1}\tl^2 \right) \tl^2.
\ee
At separations small compared with the linear dimension of the membrane, viz., $\ell_0 < \sqrt{3c}(\boltzmann T/4\stiff)^{1/4}\sqrt{S}$, we obtain
\be
\sigma^2 = \frac{1}{\bing{12}c}\sqrt{\frac{\boltzmann T}{\stiff}}\ell_0^2
\label{eq:s_steric}
\ee
This formula has the same scaling dependence on separation as the one originally postulated by Helfrich, but here we have derived the scaling dependence rather than postulated it. From our result we see that the fluctuations become small at low temperature and/or large membrane curvature modulus. 
On the other hand, for two membranes that are more widely separated apart than the linear dimension of either membrane, viz., $\ell_0 > \sqrt{3c}(\boltzmann T/4\stiff)^{1/4}\sqrt{S}$, there is a crossover to the behavior of a single, free membrane:
\be
\sigma^2 = \frac{\boltzmann TS}{\bing{4}\pi \stiff}
\label{eq:sigma_saturates}
\ee 
At large separations, each membrane would behave as if there is no hard wall potential present, and the mean square fluctuation of each membrane is then set by its total area. 

The free energy per unit area $f_s$ for a given average separation $\ell_0$ and external osmotic pressure $P$ is given by $f_s = -T\ln Z$. The functional integration over the fluctuation modes $\delta\ell_\Qv$ in Fourier space is Gaussian and \bing{yields} 
\be
f_s = P\ell_0 + \frac{3\boltzmann T c^2}{8\ell_0^2} + \frac{\boltzmann T}{\bing{2}} \!\!\int\!\!\frac{d^2Q}{(2\pi)^2}\ln\left\{ 1 + \frac{9\boltzmann T c^2}{4\stiff \ell_0^4 Q^4} \right\}.
\ee
In the above calculation, we have subtracted off a constant background contribution to $f_s$ coming from $\ell_0 \rightarrow \infty$~\cite{safran}. Taking the upper bound of the momentum integral to be $\infty$ and the lower bound to be $1/\sqrt{S}$, we obtain a \bing{closed-form} expression for $f_s$:
\ba
f_s &=& P\ell_0 + \frac{3\boltzmann T c^2}{8\ell_0^2} 
-\frac{\boltzmann T}{\bing{8}\pi S}\ln\left\{ 1+\frac{9\boltzmann T c^2 S^2}{4\stiff \ell_0^4} \right\} 
\nonumber\\
&&+ \frac{\boltzmann T}{\bing{4}\pi} \sqrt{\frac{9\boltzmann T c^2}{4\stiff \ell_0^4}}
\bigg\{
\tan^{-1}\bigg( 1-\frac{\sqrt{2}(4\stiff)^{1/4}\ell_0}{(9\boltzmann T c^2)^{1/4}\sqrt{S}} \bigg)
\nonumber\\
&&\qquad+
\tan^{-1}\bigg( 1+\frac{\sqrt{2}(4\stiff)^{1/4}\ell_0}{(9\boltzmann T c^2)^{1/4}\sqrt{S}} \bigg)
\bigg\}.
\label{eq:fs_exact}
\ea
We can consider the far- and near-field behavior of $f_s$. 
The far-field regime is described by $\ell_0 \gg \sqrt{3c}(\boltzmann T/4\stiff)^{1/4}\sqrt{S}$. To order $S^3 \ell_0^{-8}$, $f_s$ is given by
\be
f_s \approx P\ell_0 + \frac{3\boltzmann T c^2}{8\ell_0^2} 
+ \frac{9(\boltzmann T)^2 c^2 S}{\bing{32} \pi \stiff \ell_0^4} 
+ \frac{27(\boltzmann T)^3 c^4 S^3}{\bing{128} \pi \stiff^2 \ell_0^8}
\label{eq:helf-like_farfield}
\ee
In the near-field regime, $\ell_0 \ll \sqrt{3c}(\boltzmann T/4\stiff)^{1/4}\sqrt{S}$. To order $S^{-1}$, $f_s$ is given by
\ba
f_s &\approx& \frac{3\boltzmann Tc}{\bing{16} \ell_0^2} \sqrt{\frac{\boltzmann T}{\stiff}} + \frac{3\boltzmann T c^2}{8\ell_0^2}  + P\ell_0 
\nonumber\\
&&- \frac{\boltzmann T}{\bing{4} \pi S}\left\{ 1+ \ln\left( \frac{3c\sqrt{\boltzmann T} S}{2\sqrt{\stiff}\ell_0^2} \right)\right\}.
\label{eq:helf-like}
\ea
Equations~(\ref{eq:helf-like_farfield}) and (\ref{eq:helf-like}) indicate that the steric potential has two contributions: one that comes from thermal fluctuations of the \emph{zero mode} (i.e., $\ell_0$), represented by the second term in Eq.~(\ref{eq:helf-like_farfield}) and in Eq.~(\ref{eq:helf-like}), and another that is induced by thermal \emph{bending} fluctuations of the membrane (i.e., $\delta\ell(\xvp)$), represented by the third and fourth terms in Eq.~(\ref{eq:helf-like_farfield}) and the first and last terms in Eq.~(\ref{eq:helf-like}). Note that as $\ell_0$ is varied from small to large values, the \emph{bending} fluctuation contribution to the steric potential changes its scaling form from $\ell_0^{-2}$ to $\ell_0^{-4}$, and is thus much weaker at large separations. 
%Although at this point we will not delve into any detail, we note that our far-field asymptotic result has implications for studies on the order of the unbinding transition of membranes whose interactions have a fluctuation-induced steric potential $V_{{\rm fl}}$ and a direct potential $V_{{\rm DI}}$ with an attractive long tail $V_{{\rm DA}}$. In FRG-based studies (see e.g. Ref.~\cite{lipowsky1990}), the system is classified as being in the weak, intermediate or strong fluctuation regime if (for large separations $\ell_0$) $V_{{\rm DA}} \gg \ell_0^{-2}$, $V_{{\rm DI}} \sim \ell_0^{-2}$ or $V_{{\rm DA}} \ll \ell_0^{-2}$ respectively. The ``baseline" scaling form $\ell_0^{-2}$ comes from the fact that such studies have been based on Ref.~\cite{helfrich} where it was found that $V_{{\rm fl}} \sim \ell_0^{-2}$; however this result applies strictly only in the near-field regime. For large separations we should replace the ``baseline" scaling form by $\ell_0^{-4}$ (which would imply that two-dimensional fluid membranes interacting in three-dimensional space via fluctuation steric and van der Waals forces are in the \emph{intermediate} fluctuation regime rather than the strong fluctuation regime, contrary to the findings of Refs.~\cite{lipowsky1,lipowsky2}). 

In the near-field regime, to zeroth order in $S^{-1}$, we can compare our result above with the picture of Ref.~\cite{helfrich}, where the free energy cost of thermal bending fluctuations of steric membranes is described via the term $V_H = c_{{\rm fl}}(\boltzmann T)^2/\stiff \ell_0^2$; in keeping with the literature we call $V_H$ the Helfrich interaction. In the Helfrich interaction term, $c_{{\rm fl}}$ is regarded as a universal number (see, e.g., Refs.~\cite{helfrich,david}). One may regard the first term of Eq.~(\ref{eq:helf-like}) as being the analogue of the Helfrich interaction~\cite{footnote:c_fl}. However, one should note an important distinction: for $\stiff \rightarrow\infty$, $V_H$ goes to zero, and the Helfrich term thus does not address the case of osmotic pressure generated by longitudinally fluctuating flat membranes (i.e., zero mode fluctuations) 
%On the other hand, such a case is addressed by the zero mode contribution to Eq.~(\ref{eq:helf-like}). Furthermore, at zero temperature the impenetrability of hard surfaces will exist even as the steric effects generated by thermal bending fluctuations die out; this dual aspect is reflected in our Boltzmann factor $e^{-\beta f}$ not depending on the term linear in $c$ but only on the term proportional to $c^2$ in Eq.~(\ref{eq:helf-like}) at $T=0$. 
and therefore describes only the bending fluctuation ``decoration" about a flat membrane, but not the longitudinal fluctuation of the flat membrane itself.  

\begin{figure}
	\centering
		\includegraphics[width=0.44\textwidth]{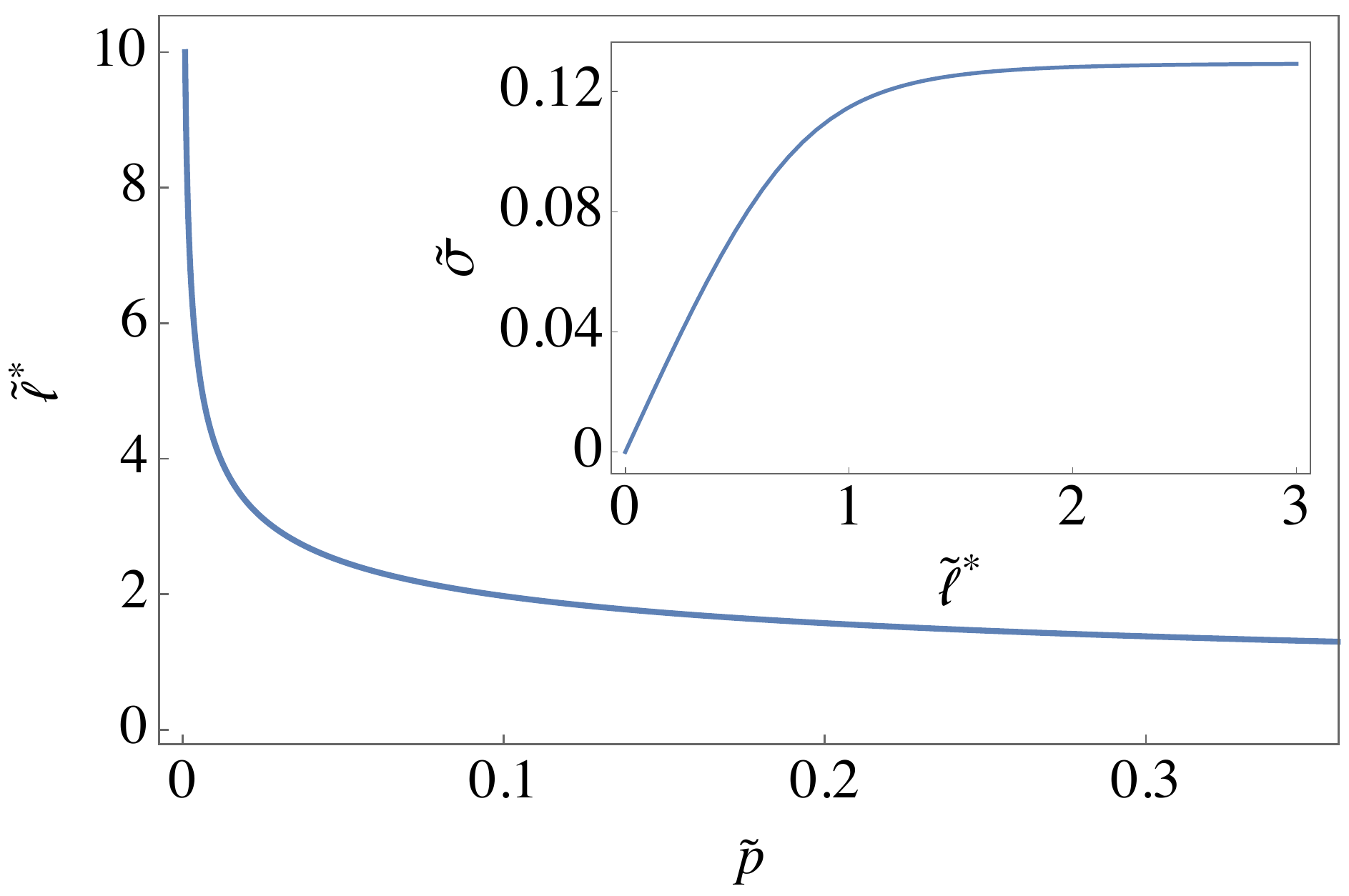}
	\caption{Steric interaction: behavior of the dimensionless separation $\tls$ as a function of dimensionless osmotic pressure $\tp$ and (inset) behavior of dimensionless rms fluctuation amplitude $\widetilde{\sigma}$ as a function of $\tls$, where $\tls \approx \langle \ell_0 \rangle/d^\ast$ is the equilibrium separation in the saddle-point approximation, $\widetilde{\sigma} \equiv \sigma/d^\ast$, $\tp \equiv \beta P (d^\ast)^3$, and $d^\ast \equiv \sqrt{3c}(\boltzmann T)^{1/4}\sqrt{S}/(4\stiff)^{1/4}$.}
\label{fig:steric_full}
\end{figure}

As the intervening region between the membranes is free to exchange volume with the surrounding reservoir whilst being subject to a constant external osmotic pressure $P$, the system behavior is best described using a constant osmotic pressure ensemble, where the equilibrium separation $\langle \ell_0 \rangle$ is determined according to 
\be
\langle \ell_0 \rangle 
\equiv \frac{1}{Z} \! \int_0^\infty \!\!\! d\ell_0 \, \ell_0 \, 
e^{-\beta S f_s}
%\nonumber\\
= -\frac{T}{S}\frac{\partial \ln Z}{\partial P}.
\ee
The equilibrium separation can be estimated in the saddle-point approximation by looking for the minimum of $f_s$ with respect to variations in $\ell_0$. Defining the dimensionless variables $\tl \equiv \ell_0/d^\ast$, $\tp \equiv \beta P (d^\ast)^3$, $\tS\equiv S/(d^\ast)^2$ and $\alpha \equiv 3c\sqrt{\boltzmann T}/8\pi\sqrt{\stiff}$, where $d^\ast \equiv \sqrt{3c}(\boltzmann T)^{1/4}\sqrt{S}/(4\stiff)^{1/4}$, we can recast Eq.~(\ref{eq:fs_exact}) in dimensionless form: 
\ba
\widetilde{f}_s &\equiv& \beta (d^\ast)^2 f_s 
\\
&=& \tp \tl + (3c^2/8)\tl^{-2} - \frac{\alpha}{\bing{2}} \ln\big( 1+\tl^{-4} \big) 
\nonumber\\
&&+ \alpha \tl^{-2}\big( \tan^{-1}(1-\sqrt{2}\tl) + \tan^{-1}(1+\sqrt{2}\tl) \big)
\nonumber
\ea
The stationarity condition $\partial \widetilde{f}_s/\partial\tl = 0$ then leads to 
\be
\tp = \frac{\frac{3c^2}{4} + \bing{2} \alpha\big( \tan^{-1}(1-\sqrt{2}\tls) + \tan^{-1}(1+\sqrt{2}\tls) \big)}{(\tls)^3}, 
\ee
where $\tls$ is the saddle-point approximation to the equilibrium separation $\langle\tl\rangle$. 
In the near-field limit ($\tls \ll 1$), we obtain to leading order
\be
\tp \approx \frac{3c}{4}\left( \bing{\frac{1}{2}} \sqrt{\frac{\boltzmann T}{\stiff}} + c \right) (\tls)^{-3}
\ee
Restoring dimensions, we have
\be
\langle \ell_0 \rangle \approx (\boltzmann T)^{1/3} \left( \frac{3c}{4}\left( \bing{\frac{1}{2}} \sqrt{\frac{\boltzmann T}{\stiff}} + c \right) \right)^{1/3} P^{-1/3}.
\ee
Thus, the equilibrium separation diverges as $P^{-1/3}$ as $P\rightarrow 0$, and vanishes as $T\rightarrow 0$ (the pair of membranes simply collapse onto each other at zero temperature) as one would expect. 

In the far-field regime ($\tls \gg 1$), we have to the order of $(\tls)^{-5}$ 
\be
\tp \approx \frac{3c^2}{4(\tls)^{3}} + \frac{3c\sqrt{\boltzmann T/\stiff}}{\bing{4}\pi (\tls)^5}. 
\ee
The equilibrium separation increases monotonically with decreasing osmotic pressure. 

The behaviors of $\tls$ and $\widetilde{\sigma}$ are plotted in Fig.~\ref{fig:steric_full}. The rms fluctuation amplitude vanishes at zero inter-membrane separation (reflecting the hard-wall constraint), scales linearly with the inter-membrane separation for small values of the separation, and saturates at a constant value set by the cross-sectional area of the membrane for large values of the separation.

\section{Feynman-Kleinert variational approximation}
\label{sec:f-k}

In the previous section we derived an effective Hamiltonian [Eq.~(\ref{eq:H_eff})] which accounts for the steric constraint via the hard potential, and applied it to study a membrane interacting with a hard wall via only Helfrich and steric forces. We were able to obtain \bing{closed-form} expressions for the rms fluctuation amplitude and free energy per unit area of the membrane [cf. Eqs.~(\ref{eq:sigma_exact}) and (\ref{eq:fs_exact})] because for such interactions the form of the effective Hamiltonian is Gaussian. In realistic systems the soft potential rarely has a Gaussian form, and further approximations will have to be made on the partition function. One such approximation \bing{is} the Feynman-Kleinert (FK) variational approximation~\cite{feynman_kleinert}, developed originally for the quantum-mechanical partition function of an anharmonic oscillator, but can equally well be applied to the classical partition function of thermally fluctuating membranes~\cite{podgornik_parsegian}. 

The application of the variational approximation~\cite{feynman} begins with a trial partition function $Z_1$ that is Gaussian in fluctuations $\delta\ell$, viz., 
\be
Z_1 \! = \! \int_{0}^{\infty} \!\!\!\! d\ell_0 \!\! \prod_{\{ \xvp \}} \! \int_{-\infty}^{\infty} \!\! d\delta\ell(\xvp)
e^{-\beta \! \int \! d^2 \xp \big( \frac{\stiff}{2} (\np^2 \ell(\xvp))^2 + V_{{\rm trial}} \big)}
\label{eq:Z1_unconstrained}
\ee
where 
\be
V_{{\rm trial}} = \frac{1}{2}B\,(\delta\ell(\xvp))^2 + w_0(\ell_0) + P \, \ell(\xvp).
\label{eq:Vtrial}
\ee
The variational principle relies on Jensens' inequality, viz.,
\be
Z > e^{-\beta\langle V - V_{{\rm trial}} \rangle_1} Z_1, 
\ee
where the notation $\langle \dots \rangle_1$ denotes Boltzmann averaging with respect to the statistics of both $\ell_0$ and $\{\delta\ell(\xvp)\}$ specified by $Z_1$. 
The best estimate is obtained by looking for the maximum upper bound on the right hand side. There are two unknowns: (i)~an unknown parameter $B$, which is related to the mean square fluctuation of the membrane, and (ii)~an unknown function $w_0$. The form of $w_0$ is fixed in terms of $B$ by requiring that 
\be
\langle V - V_{{\rm trial}} \rangle_1 = 0.
\label{eq:jensens}
\ee
The unknown parameter $B$ is then determined by optimizing $Z_1$.  

To see how the FK variational approximation works in the context of fluctuating membranes, let us define a restricted trial partition function $Z_{\sigma}$, viz., 
\be
Z_{\sigma} \equiv  
\prod_{\{ \xvp \}} \! \int_{-\infty}^{\infty} \! d\delta\ell(\xvp) 
e^{-\beta \! \int \! d^2 \xp ( \frac{\stiff}{2} (\np^2 \ell(\xvp))^2 + V_{{\rm trial}} )},
\ee
and the associated average $\langle \dots \rangle_{\sigma}$, viz., 
\begin{widetext}
\be
\langle \dots \rangle_{\sigma} \equiv \frac{\prod_{\{ \xvp \}}\int_{-\infty}^{\infty} d\delta\ell(\xvp) \, (\dots) \,
e^{-\beta \int d^2 \xp ( \frac{\stiff}{2} (\np^2 \ell(\xvp))^2 + V_{{\rm trial}} )}
}{\prod_{\{ \xvp \}}\int_{-\infty}^{\infty} d\delta\ell(\xvp) 
e^{-\beta \int d^2 \xp ( \frac{\stiff}{2} (\np^2 \ell(\xvp))^2 + V_{{\rm trial}} )}}.
\ee
By using Eqs.~(\ref{eq:V_eff}), (\ref{eq:Vtrial}) and (\ref{eq:jensens}), we obtain
\ba
&&Z_{\sigma} \prod_{\{ \xvp \}}\int_{-\infty}^{\infty} d\delta\ell(\xvp) \, ( V - V_{{\rm trial}} ) \,
e^{-\beta \int d^2 \xp ( \frac{\stiff}{2} (\np^2 \ell(\xvp))^2 + V_{{\rm trial}} )} = 0
\nonumber\\
&&\Rightarrow w_0(\ell_0) = w_{\sigma^2} (\ell_0) - \frac{B}{2}\sigma^2,
\label{eq:jensens2}
\ea
where the variational estimates of the mean square fluctuation $\sigma^2$ and the interaction energy $w_{\sigma^2}$ are defined by 
\begin{subequations}
\ba
&&\sigma^2 \equiv \langle (\delta\ell(\xvp))^2 \rangle_{\sigma};
\\
&&w_{\sigma^2} (\ell_0) \equiv \langle w(\ell) \rangle_{\sigma}.
\label{eq:wvar}
\ea
\end{subequations}
We compute the mean square fluctuation: 
\ba
\sigma^2 &=& \int \frac{d^2\xp}{S} \langle (\delta\ell(\xvp))^2 \rangle_{\sigma} = 
\frac{\bing{S^{-1}} \!\! \int \!\! \frac{d^2 Q}{(2\pi)^2} 
\prod_{\{ \Qv \bing{> \bm{0}} \}}\int d\delta\ell_{\Qv}^{{\rm re}} d\delta\ell_{\Qv}^{{\rm im}}
\big( (\delta\ell_Q^{{\rm re}})^2 + (\delta\ell_Q^{{\rm im}})^2 \big) 
e^{-\bing{\beta S^{-1} \!\!\! \sum\limits_{\Qv > \bm{0}}}
(\stiff Q^4 + B)\big( (\delta\ell_Q^{{\rm re}})^2 + (\delta\ell_Q^{{\rm im}})^2 \big)}}{\prod_{\{ \Qv \bing{> \bm{0}} \}}\int d\delta\ell_{\Qv}^{{\rm re}} d\delta\ell_{\Qv}^{{\rm im}}
e^{-\bing{\beta S^{-1} \!\!\! \sum\limits_{\Qv > \bm{0}}} 
(\stiff Q^4 + B)\big( (\delta\ell_Q^{{\rm re}})^2 + (\delta\ell_Q^{{\rm im}})^2 \big)}}
%\nonumber\\
%&=& 
%\int \frac{d^2 Q}{(2\pi)^2} \left( \frac{\bing{\boltzmann T}}{\stiff Q^4 + B} \right) 
%= \int_{\frac{1}{\sqrt{S}}}^{\infty} \! \frac{dQ}{2\pi} \left( \frac{\bing{\boltzmann TQ}}{\stiff Q^4+B} \right) 
\nonumber\\
&=& 
\frac{\boltzmann T}{\bing{8}\sqrt{\stiff B}} \left( 1 - \frac{2}{\pi}\tan^{-1}\left( \frac{1}{S}\sqrt{\frac{\stiff}{B}} \right) \right)
\label{eq:ms_fluct_exact}
\ea
\end{widetext}
\bing{In the above calculation, we have made use of the continuum representation of the wave-vector sum,  
$S^{-1} \!\! \sum\limits_{ \{ \Qv \} } \rightarrow \int \!\! \frac{d^2Q}{(2\pi)^2}$, and restricted the wave-vector sum to those wave-vectors that are positive, the reason being that $\delta\ell(\xvp)$ is real and thus the components $\delta\ell_{\Qv}^{{\rm re}}$ (and $\delta\ell_{\Qv}^{{\rm im}}$) are not independent: $\delta\ell_{\Qv}^{{\rm re}} = \delta\ell_{-\Qv}^{{\rm re}}$ and $\delta\ell_\Qv^{{\rm im}} = -\delta\ell_{-\Qv}^{{\rm im}}$. 
In the second line of Eq.~(\ref{eq:ms_fluct_exact}),} 
we have included a correction term due to the finite size of the membrane. In the limit of an infinitely large membrane (which effectively means that the square root of the membrane's projected cross-sectional area is much greater than the inter-membrane separation), the above result simplifies to 
\be
\sigma^2 =  \frac{\boltzmann T}{\bing{8}\sqrt{\stiff B}}
\label{eq:sigmaC}
\ee
and
\be
B = \frac{(\boltzmann T)^2}{\bing{64}\stiff\sigma^4},
\label{eq:sigmaB}
\ee
\bing{which agrees with Eq.~(10) in Ref.~\cite{podgornik_parsegian}.} 
Let us compute $w_{\sigma^2}(\ell_0)$ by taking explicitly into account the Gaussian variational {\sl Ansatz}, leading to
\ba
w_{\sigma^2}(\ell_0) &\equiv& \langle w(\ell(\xvp)) \rangle_\sigma = \int_{-\infty}^{\infty}\!\frac{dk}{2\pi} w(k) \langle e^{ik\ell(\xvp)} \rangle_\sigma
\nonumber\\
%&\approx& \int_{-\infty}^{\infty}\!\frac{dk}{2\pi} w(k) \, e^{ik\ell_0 - \frac{1}{2}k^2\sigma^2}
%\nonumber\\
&=& \int_{-\infty}^{\infty}\!\frac{dk}{2\pi} \!\! \int_{-\infty}^{\infty}\!\!\!d\ell \, w(\ell)\, e^{-ik(\ell-\ell_0)- \frac{1}{2}k^2\sigma^2}
\nonumber\\
&=& \int_{-\infty}^{\infty}\!\!\!\frac{d\ell}{\sqrt{2\pi\sigma^2}} w(\ell)\,e^{-\frac{(\ell-\ell_0)^2}{2\sigma^2}}. 
\label{eq:w_var}
\ea
In the second line, we have performed a second order cumulant expansion for the Gaussian variational {\sl Ansatz}, followed by an inverse Fourier transform of $w$, and $\ell$ is a dummy variable [not to be confused with the thermally fluctuating \emph{field} $\ell(\xvp)$ of the first line] that runs from $-\infty$ to $\infty$. Next, we have integrated over $k$. The final result thus depends only on $\ell_0$ and $\sigma$. Knowing the form of the interaction potential $w(\ell)$, we can plug it into Eq.~(\ref{eq:w_var}) to obtain its variational estimate. 

In what follows, it is useful to define the variational free energy for a given $\ell_0$:  
\be
Z_1 = \int_0^\infty \!\!\!\! d\ell_0 \, e^{-\beta S f_{{\rm var}}} = \int_0^\infty \!\!\!\! d\ell_0 \, Z_{\sigma}.
\label{eq:Z_vardef}
\ee
To determine the variational free energy, we first express Eqs.~(\ref{eq:Z1_unconstrained}) and (\ref{eq:Vtrial}) in Fourier space:
\ba
Z_1 &=& \int \! d\ell_0 \prod_{\{ \Qv \bing{ > \bm{0}} \}} \! \int \! d\delta\ell_Q^{{\rm re}} \! \int \! d\delta\ell_Q^{{\rm im}}
e^{- \beta S (P \ell_0 + w_0(\ell_0)) } 
\nonumber\\
&&\quad\times e^{- \bing{ \beta S^{-1} \!\! \sum\limits_{\Qv > \bm{0}} } (\stiff Q^4 + B) |\delta\ell_Q|^2}.
\label{eq:Z1unintegrated}
\ea
Integrating over real and imaginary modes of $\delta\ell_Q$, we obtain
\ba
&&Z_{\sigma} = 
\prod_{\{ \Qv \bing{ > \bm{0} } \}} 
\int \! d\delta\ell_Q^{{\rm re}} 
\int \! d\delta\ell_Q^{{\rm im}} 
e^{- \bing{ S^{-1} \!\! \sum\limits_{\Qv > \bm{0}} } (\stiff Q^4 + B) |\delta\ell_Q|^2}
\nonumber\\
&&= \prod_{\{ \Qv \bing{ > \bm{0}} \}} \left( \frac{ \bing{\pi \boltzmann T} }{\stiff Q^4 + B} \right) = 
e^{\bing{ \frac{1}{2} \!\! \sum\limits_{\Qv > \bm{0}} } \ln \left( \frac{\bing{\pi \boltzmann T}}{\stiff Q^4 + B} \right)} 
\nonumber\\
%&&= 
%e^{S \! \int \! \frac{dQ}{2\pi} Q\ln\left( \frac{2 \pi \boltzmann T}{\stiff Q^4 + B} \right)}
%\nonumber\\
&&\xrightarrow{regularise} e^{S \! \int \! \frac{dQ}{\bing{4}\pi} Q\ln\left( \frac{Q^4}{Q^4 + (B/\stiff)} \right)} = 
e^{-\frac{S}{\bing{8}}\sqrt{\frac{B}{\stiff}}} 
\label{eq:Z_tr}
\ea
The regularisation subtracts off the constant (and divergent) contribution of membranes that are infinitely far apart~\cite{safran}. 
Equation (\ref{eq:Z1unintegrated}) thus becomes 
\be
Z_1 = \int \! d\ell_0 \, 
e^{- \beta S (P \ell_0 + w_0(\ell_0) + \frac{\boltzmann T}{\bing{8}}\sqrt{\frac{B}{\stiff}}) }.
\label{eq:Z1_first}
\ee
%Equation~(\ref{eq:Z1_first}) can also be derived in another way; see App.~\ref{app:Z1_derive}. 
Substituting for $w_0(\ell_0)$ its value from Eq.~(\ref{eq:jensens2}) and using Eqs.~(\ref{eq:H_eff}) and (\ref{eq:wvar}), we obtain 
\be
Z_1 \! = \! \int \! d\ell_0 
e^{-\beta S \left( \tw_{\sigma^2} + \frac{9 \boltzmann T c^2 \sigma^2 }{8 \ell_0^4} + 
 \frac{3 \boltzmann T c^2}{8 \ell_0^2}  + P \ell_0 - \frac{1}{2} B \sigma^2 + 
\frac{\boltzmann T\sqrt{B}}{\bing{8}\sqrt{\stiff}} \right) },
\ee
or equivalently, from Eqs.~(\ref{eq:sigmaC}) and (\ref{eq:sigmaB}), 
\be
Z_1 \! = \! \int \! d\ell_0 \, 
e^{- \beta S \left( \tw_{\sigma^2} + \frac{(\boltzmann T)^2}{\bing{128} \stiff \sigma^2} + \frac{9 \boltzmann T c^2 \sigma^2 }{8 \ell_0^4} + 
 \frac{3 \boltzmann T c^2}{8 \ell_0^2} + P\ell_0 \right) }.
\ee
Comparing with Eq.~(\ref{eq:Z_vardef}), we see that the variational free energy per unit area is given by 
\be
f_{{\rm var}} = \frac{(\boltzmann T)^2}{\bing{128} \stiff \sigma^2} + \frac{9 \boltzmann T c^2 \sigma^2 }{8 \ell_0^4} + 
 \frac{3 \boltzmann T c^2}{8 \ell_0^2} + \tw_{\sigma^2} + P\ell_0.
\label{eq:varF}
\ee
In the language of Ref.~\cite{lipowsky1990}, the first three terms can be regarded as comprising a repulsive fluctuation potential. The fourth term is a fluctuation-renormalized version of the direct potential, and thus $f_{{\rm var}}$ {\em cannot} be simply regarded as a sum of fluctuation and direct potentials. The additive sum would be valid in the so-called weak fluctuation regime, defined by Ref.~\cite{lipowsky1990} to be one in which the attractive tail of the direct potential is stronger than the repulsive fluctuation potential. 

On the other hand, $f_{{\rm var}}$ is not applicable to the strong fluctuation regime, defined to be one in which the fluctuation potential is stronger than the attractive tail of the direct potential, because in such a regime the fluctuations are strongly nonlinear and the Gaussian approximation assumed by the VGA is no longer reliable. The VGA-based $f_{{\rm var}}$ thus applies to a regime intermediate between the weak and strong fluctuation regimes, which as already stated, is exactly the regime we want to address. 
%Such a free energy seems appropriate for the intermediate fluctuation regime, defined in Ref.~\cite{lipowsky1990} to be the regime where the attractive tail of the direct potential is comparable in strength to the tail of the repulsive fluctuation potential. In the weak fluctuation regime, where the former is stronger than the latter, simple additivity of fluctuation and direct potentials is legitimate; in the strong fluctuation regime, where the tail of the fluctuation potential is stronger than the attractive tail of the direct potential, the Gaussian approximation to the nonlinearities of bending fluctuations assumed by the VGA is not valid. 

To obtain a relation between $\sigma^2$ and $\ell_0$, we vary $f_{{\rm var}}$ with respect to $\sigma^2$; this yields  
\be
\frac{\partial \tw_{\sigma^2}}{\partial \sigma^2} -
\frac{(\boltzmann T)^2}{\bing{128}\stiff \sigma^4} + \frac{9 \boltzmann T c^2}{8 \ell_0^4} = 0     
\label{eq:implicit_relation}
\ee 
This can be rewritten in the form 
\be
\frac{\partial \tw_{\sigma^2}}{\partial \sigma^2} = \frac{1}{2} B - \frac{9 \boltzmann T c^2}{8 \ell_0^4}, 
\label{eq:niceform}
\ee
which is similar to Eq.~(17) of Ref.~\cite{podgornik_parsegian}, with the extra contribution coming from the steric potential. 

The equilibrium value of $\ell_0$ for a given external osmotic pressure $P$ is defined by 
\be
\langle \ell_0 \rangle_1 
%= -\frac{T}{S}\frac{\ln Z_1(P)}{\partial P} 
= \frac{1}{Z_1(P)} \! \int_0^\infty \!\!\! d\ell_0 \, \ell_0 \, 
e^{-\beta S f_{{\rm var}}(P) }
\label{eq:equilibrium_ell0}
\ee
The equilibrium separation $\langle \ell_0 \rangle_1$ is thus a function of $P$. We invert the above relation to determine how $P$ depends on $\langle \ell_0 \rangle_1$, which gives the equation of state. For this one generally has to resort to numerical means. On the other hand, one can approximate $\langle \ell_0 \rangle_1$ by its saddle-point value $\ell_0^\ast$, obtained by minimizing $f_{{\rm var}}$ over $\ell_0$:
\be
\left.\frac{\partial f_{{\rm var}}}{\partial \ell_0}\right\vert_{\ell_0^\ast} = 0. 
\label{eq:sp_approximation}
\ee
This relation yields $\ell_0^\ast$ as a function of the external osmotic pressure $P$. To summarize: in the Feynman-Kleinert approach, one decomposes the thermally fluctuating field into two types of contributions, a zero-mode contribution $\ell_0$ and a finite wave-vector contribution $\{ \delta\ell_\Qv\}_{\Qv\neq 0}$, approximates $\sigma^2$ by means of a Gaussian kernel but retains the full nonlinear dependence of the free energy on $\ell_0$ (present in $\widetilde{w}_{\sigma^2}$), which is then minimized over $\ell_0$. This leads to an improved accuracy over that of the more conventional form of VGA~\cite{feynman}, which approximates both the thermal equilibrium average of an observable and its mean square fluctuation by means of a Gaussian kernel.

\subsection{Applications}
In what follows, we apply our formalism and the variational approach developed in previous sections to study the physical behavior of four different model membrane systems in the regime where $\ell_0 \ll \sqrt{S}$. 

First, we analyze a system consisting of equally charged impenetrable membranes, with an intervening solution of multivalent counterions, assuming that the surface charge density is sufficiently large so that the system is in the strong coupling (SC) regime of electrostatics~\cite{netz-sc,moreira-netz,boroudjerdi-netz,kanduc-thesis}. In this regime, for the case of two planar, infinitely rigid membranes, the osmotic  pressure decomposes into a simple sum of two contributions: one describing the entropy reduction of each counterion and another describing the electrostatic interaction between the counterion and a charged membrane. We will call the interaction potential of such a system the {\em Moreira-Netz potential}.   

Our second system has for its interaction potential an {\em attractive square well}, which is a model for describing the adhesion of membranes by short-range attractive forces, studied e.g. in Ref.~\cite{netz_unbinding}. 

In our third system, the direct interaction is modeled by a {\em Morse potential} which has been used in studies of biological membrane systems and interactions of DNA molecules in solution~\cite{todd,hoang,manghi}. The Morse potential consists of a repulsive exponential term which mimics the hydration and excluded volume repulsion of the membranes and an attractive exponential term (with twice the decay width of the repulsive term) which mimics the attractive long-raneg tail. 

Finally, we consider a system consisting of two membranes interacting via a generic soft potential consisting of a {\em hydration term and a van der Waals term} approximated in the non-retarded planar limit~\cite{wennerstrom,petrache}.

\subsubsection{Moreira-Netz potential}
Let us first consider the case of a pair of strongly charged membranes in solution with counterions of valence $q$ in the intervening region, of charge sign opposite to the charge on the membranes. As the membranes are strongly charged, the system is in what is known as the strong coupling (SC) regime~\cite{netz-sc,moreira-netz,boroudjerdi-netz,kanduc-thesis}, where Poisson-Boltzmann (PB) theory and its fluctuation-corrected version break down. The SC regime is characterized by $\Xi \gtrsim 1$, where the coupling strength $\Xi \equiv 2\pi q^3 \bjerrum^2 \sigma_s$. Here, $\sigma_s$ is the number density of charges on the surface of either membrane, and $\bjerrum \equiv e^2/4\pi\epsilon\epsilon_0\boltzmann T$ is the Bjerrum length ($e$ is the elementary charge). 
The Bjerrum length provides a measure of the interaction strength between a pair of unit charges at a given temperature $T$. 
The regime for which PB theory holds is known as the weak-coupling regime, and is characterized by $\Xi \lesssim 1$. 

In the SC regime, the interaction between two equally charged hard planar surfaces is described by the Moreira-Netz free energy per unit area~\cite{netz-sc,kanduc-thesis}:
\be
f_{{\rm hard}} = 4\pi\boltzmann T \bjerrum \sigma_s^2 \left( \frac{\ell}{2\gouy} - \ln \frac{\ell}{2\gouy}  \right),
\label{eq:fsc_hard}
\ee
where $\gouy\equiv 1/2\pi q \bjerrum \sigma_s$ is the Gouy-Chapman length, which is inversely proportional to both the valence of the counterion species and the surface charge density of each plate. 
\bing{The first term in Eq.~(\ref{eq:fsc_hard}) reflects the electrostatic attraction between the counterions and the charged plates, whilst the second term originates from the entropic  pressure of the counterions confined between the two plates.} The Moreira-Netz free energy was derived for a system where the charged plates are held fixed and are effectively rigid.
%not allowed to equilibrate (and thus corresponds to a \emph{constant volume} ensemble). 

We can generalize the Moreira-Netz free energy to the case of two thermally fluctuating membranes, where the membranes are allowed to equilibrate under an externally applied osmotic pressure (thus we are considering a \emph{constant osmotic pressure} ensemble). We replace $\ell$ with $\ell(\xvp)$, where the inter-membrane separation now depends on the transverse coordinate $\xvp = (x,y)$. As before, we write the separation as the sum of mean and fluctuating contributions: $\ell(\xvp) = \ell_0 + \delta\ell(\xvp)$, where $\ell_0$ and $\delta\ell(\xvp)$ are thermal variables. The corresponding soft potential $\tw$ is given by 
\ba
\tw &=& -4\pi\boltzmann T\gouy\bjerrum\sigma_s^2 \left( \ln \frac{\ell(\xvp)}{2\gouy} - \frac{\ell(\xvp)}{2\gouy} \right)
\nonumber\\
&=& -4\pi\boltzmann T\gouy\bjerrum\sigma_s^2 \bigg( \ln \frac{\ell_0}{2\gouy} + \frac{\delta\ell(\xvp)}{\ell_0}  - \frac{(\delta\ell(\xvp))^2}{2\ell_0^2} 
\nonumber\\
&&\quad- \frac{\ell_0}{2\gouy} -  \frac{\delta\ell(\xvp)}{2\gouy} \bigg)
\ea
In the second step, we have Taylor expanded the logarithm to quadratic order in $\delta\ell(\xvp)$. 
On applying the Gaussian approximation of Eq.~(\ref{eq:w_var}), we obtain
\be
\beta\tw_{\sigma^2}(\ell_0) = -4\pi\gouy\bjerrum\sigma_s^2 \bigg( \ln \frac{\ell_0}{2\gouy} - \frac{\ell_0}{2\gouy} - \frac{\sigma^2}{2\ell_0^2} \bigg)
\ee
The variational free energy per unit area is then given by [cf. Eq.~(\ref{eq:varF})]
\ba
\beta f_{{\rm var}} &=& \frac{\boltzmann T}{\bing{128} \stiff \sigma^2} + \frac{9 c^2 \sigma^2 }{8 \ell_0^4} + 
 \frac{3 c^2}{8 \ell_0^2} + \beta P \ell_0
 \nonumber\\
 &&-4\pi\gouy\bjerrum\sigma_s^2 \bigg( \ln \frac{\ell_0}{2\gouy} - \frac{\ell_0}{2\gouy} - \frac{\sigma^2}{2\ell_0^2} \bigg)
\ea
where $P$ is the external osmotic pressure applied to keep the membranes at a constant average separation. 
%\begin{figure}
	%	\includegraphics[width=0.4\textwidth]{sigma_hard_vs_soft.pdf}
	%\caption{Behavior of the rescaled rms fluctuation amplitude $\sigma/\gouy$ [defined in Eq.~(\ref{eq:sigma_hard_vs_soft})] with $t = 0.1$, $c = 1$ and $\widetilde{\sigma}_s = 1$, studied for the three cases: (i)~$q=1$ (blue), (ii)~$q=2$ (green dashed), and (iii)~$q=3$ (red dot-dashed). For comparison, we have shown the rms fluctuation amplitude of a membrane interacting only sterically [black; cf. Eq.~(\ref{eq:s_steric})].}
%\label{fig:sigma_hard_vs_soft}
%\end{figure}
Varying $f_{{\rm var}}$ with respect to $\sigma$, we obtain 
\be
\sigma^2 = \left( 1 + \frac{8\sigma_s \ell_0^2}{9 c^2 q} \right)^{-1/2} \sqrt{\frac{\boltzmann T}{\stiff}} \frac{\ell_0^2}{\bing{12}c}.
\ee
To form dimensionless variables, let us rescale $\sigma$ and $\ell_0$ in units of $\mu_0 \equiv q \mu$ (since the Gouy-Chapman length $\gouy$ varies as $q^{-1}$; we are interested in the effect of varying the valence $q$, so the basic lengthscale should be independent of $q$). Let us also define dimensionless variables $t \equiv \boltzmann T/\stiff$, $\widetilde{\ell}_0  \equiv \ell_0/\mu_0$, $\widetilde{\sigma}_s \equiv \mu_0^2 \sigma_s $ and $\widetilde{\sigma} \equiv \sigma/\mu_0$, whence the above equation can be put in the form
\be
\widetilde{\sigma}^2 = \frac{\sqrt{t} \, \widetilde{\ell}_0^2}{\bing{12}c} \left( 1 + \frac{8\widetilde{\sigma}_s \widetilde{\ell}_0^2}{9 c^2 q} \right)^{-1/2}.
\label{eq:sigma_hard_vs_soft}
\ee
This formula holds strictly for the SC regime where $\Xi \gg 1$. 
%Its behavior is shown in Fig.~\ref{fig:sigma_hard_vs_soft}. 
%We see that $\sigma$ is larger for smaller $\Xi$, and approaches the value for a pair of membranes interacting purely via steric forces for $\Xi\rightarrow\infty$.  
Substituting the value of $\sigma^2$ into our expression for $f_{{\rm var}}$ yields $f_{{\rm soft}}$, the effective interaction energy between two ``soft" surfaces: 
\ba
 \label{eq:fsc_soft}
 \beta f_{{\rm soft}} &\!\!=\!\!& \frac{3c}{\bing{16}\ell_0^2}\sqrt{\frac{\boltzmann T}{\stiff}} 
\left( 1 + \frac{8\sigma_s \ell_0^2}{9c^2 q} \right)^{1/2} + \frac{3 c^2}{8 \ell_0^2} 
\\
 &&-4\pi\gouy\bjerrum\sigma_s^2\ln\frac{\ell_0}{2\gouy} + (\beta P + 2\pi\bjerrum\sigma_s^2)\ell_0.
\nonumber
\ea
The system still needs to be equilibrated with respect to $\ell_0$.
\begin{figure}
		\includegraphics[width=0.44\textwidth]{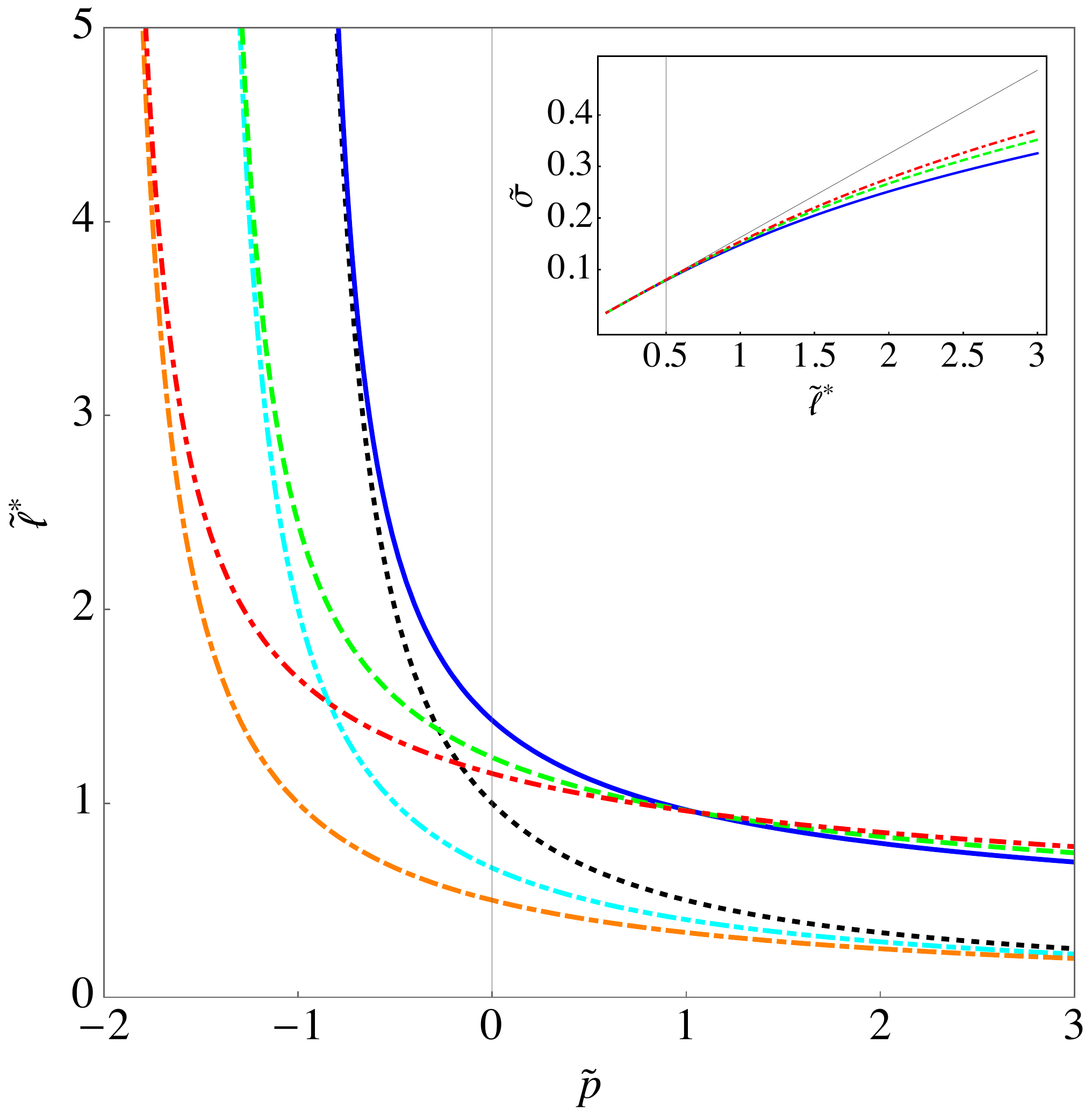}
	\caption{Moreira-Netz potential. Behavior of $\tls \equiv \ell_0^\ast/\mu_0$ (where $\ell_0^\ast$ is the saddle-point approximation to the equilibrium inter-membrane separation and $\mu_0 \equiv 1/2\pi \bjerrum \sigma_s$) as a function of rescaled osmotic pressure $\tp \equiv \beta P/4\pi\bjerrum\sigma_s^2$ [given by Eq.~(\ref{eq:p_soft})] and (inset) behavior of the rescaled rms fluctuation amplitude $\widetilde{\sigma} \equiv \sigma/\mu_0$ as a function of $\tls$ [defined in Eq.~(\ref{eq:sigma_hard_vs_soft})] for $t = 0.1$, $c = 1$ and $\widetilde{\sigma}_s = 1$, studied for the following three cases: (i)~$q=2$ (blue), (ii)~$q=3$ (green dashed); and (iii)~$q=4$ (red dot-dashed). 
For comparison, we have displayed the behavior of the disjoining osmotic pressure (horizontal axis) due to counterions between two fixed charged plates as a function of the inter-plate separation (vertical axis) for counterion valences $q=2$ (black dotted), $q=3$ (cyan dot-dot-dashed) and $q=4$ (orange dot-dashed-dashed) [obtained by differentiating Eq.~(\ref{eq:fsc_hard})], and also shown (cf. inset) the behavior of $\widetilde{\sigma}$ of a membrane interacting only sterically with a hard wall [black; cf. Eq.~(\ref{eq:s_steric})].}
\label{fig:sc}
\end{figure}
%To facilitate comparison of the behavior of the interaction energy of ``hard" (i.e., non-thermally fluctuating) charged planar surfaces $f_{{\rm hard}}$ in Eq.~(\ref{eq:fsc_hard}) with $f_{{\rm soft}}$ in Eq.~(\ref{eq:fsc_soft}) (which describes the effective interaction between ``soft" charged surfaces with the thermal undulations integrated out), we 
Let us define the dimensionless variables: $\tl  \equiv \ell_0/\mu_0$ and $\tp \equiv \beta P/4\pi\bjerrum\sigma_s^2$. 
Equation~(\ref{eq:fsc_soft}) then becomes 
\ba
\widetilde{f}_{{\rm soft}}(\tl) &=& q(\tp + \frac{1}{2})\tl  - \ln \frac{q\tl}{2} + \frac{3 q c^2}{16 \widetilde{\sigma}_s \tl^2}
\nonumber\\
&&+ \frac{3 q c \sqrt{t}}{\bing{32} \widetilde{\sigma}_s \tl^2} \sqrt{1+\frac{8 \widetilde{\sigma}_s \tl^2}{9c^2 q}},
\label{eq:f_hard_soft}
\ea
where $\widetilde{f} \equiv \beta f/4\pi\mu\bjerrum\sigma_s^2$. 
In the saddle point approximation, the equilibrium separation $\tls$ is given by the solution to the equation $\partial\widetilde{f}_{{\rm soft}}(\tl)/\partial \tl = 0$. We obtain
\ba
\tp 
&=& \frac{1}{\tls} - \frac{q}{2} + \frac{3 q c^2}{8 \widetilde{\sigma}_s (\tls)^3} 
   - \frac{\sqrt{t}}{\bing{12} c \tls \sqrt{1 + \frac{8 (\tls)^2 \widetilde{\sigma}_s}{9c^2 q}}} 
   \nonumber\\
   &&
+ \frac{3 q c \sqrt{t} \sqrt{\frac{8 (\tls)^2 \widetilde{\sigma}_s}{9 c^2 q}+1} }{\bing{16} \widetilde{\sigma}_s (\tls)^3} 
\label{eq:p_soft}
\ea
The last three terms on the RHS of Eq.~(\ref{eq:p_soft}) describe corrections to the equilibrium separation induced by the thermal fluctuations of the membrane. 

In Fig.~\ref{fig:sc}, we display a plot of $\tls$ as a function of $\tp$ for the case of thermally fluctuating membranes, where we have also plotted the case of fixed charged plates for comparison. In the inset we show the behavior of $\widetilde{\sigma} \equiv \sigma/\mu_0$ as a function of $\tls$ for thermally fluctuating membranes. The behaviors are plotted for $t = 0.1$, $c = 1$, $\widetilde{\sigma}_s = 1$, and three choices of the counterion valence: $q=2,3,4$. As we see from the figure, in the SC regime a bound state always forms at zero external osmotic pressure. Secondly, the bound state separation is %larger 
smaller for larger counterion valences. Thirdly, at zero osmotic pressure the equilibrium bound state separation of thermally fluctuating membranes is consistently enhanced relative to the corresponding separation of the hard plates for the same counterion valences. Finally, at large values of $\tls$ the pressure curves tend to a saturation value of $\tp = -q/2$ for counterion valence $q$, identical to the case with no thermal fluctuations.  Our prediction is thus that for multivalent counterion-mediated interactions, the equilibrium inter-membrane spacing for soft and rigid surfaces, everything else being the same, should differ and the larger the valency, the more they should differ. The predicted difference in equilibrium spacing can be in excess of a factor of 2, see Fig.~\ref{fig:sc}.

%\RP{Some comments: The graph should be for q = 2,3,4, for q=1 its not really SC. Maybe you should also increase the range of separation in the Fig. to see how the pressure saturates at a finite negative value, if indeed it does. In general the negative p axis should be extended.}

\subsubsection{Attractive square well potential}
\label{sec:attractive_square_well}

Following Ref.~\cite{netz_unbinding}, we describe the binding potential by an attractive square well potential, and study the effect that this square well potential has on the fluctuation and free energy behavior of a membrane using the variational  framework we developed in Sec.~\ref{sec:f-k}. As before we represent our system by a hard wall at $z=0$ and a membrane whose surface is at a separation $z=\ell(\xvp)$. The square well potential $V(z)$ is described by $V(z)=V_0$ for $0 < z \leq \width$ and $V(z) = 0$ for $z > \width$. For an attractive potential $V_0 < 0$. 
 
We represent the square well potential in terms of Heaviside functions $\Theta$, i.e.,  
\be
\tw(\ell(\xvp)) = V_0 (\Theta(\ell(\xvp)) - \Theta(\ell(\xvp) - \width)). 
\ee
The Gaussian approximation $\tw_{\sigma^2}$ is given by Eq.~(\ref{eq:w_var}): 
\ba
\tw_{\sigma^2} &\!\!=\!\!& 
 \int_{-\infty}^{\infty}\!\!\!\frac{d\ell}{\sqrt{2\pi\sigma^2}} 
 V_0 (\Theta(\ell) - \Theta(\ell - \width))
 \,
 e^{-\frac{(\ell-\ell_0)^2}{2\sigma^2}}
 \nonumber\\
 &\!\!=\!\!&
 \frac{V_0}{2}\left( \text{erf}\left( \frac{\ell_0}{\sqrt{2}\sigma} \right) - 
 \text{erf}\left( \frac{\ell_0 - \width}{\sqrt{2}\sigma} \right) \right).
\ea
\begin{figure}
	\centering
		\includegraphics[width=0.44\textwidth]{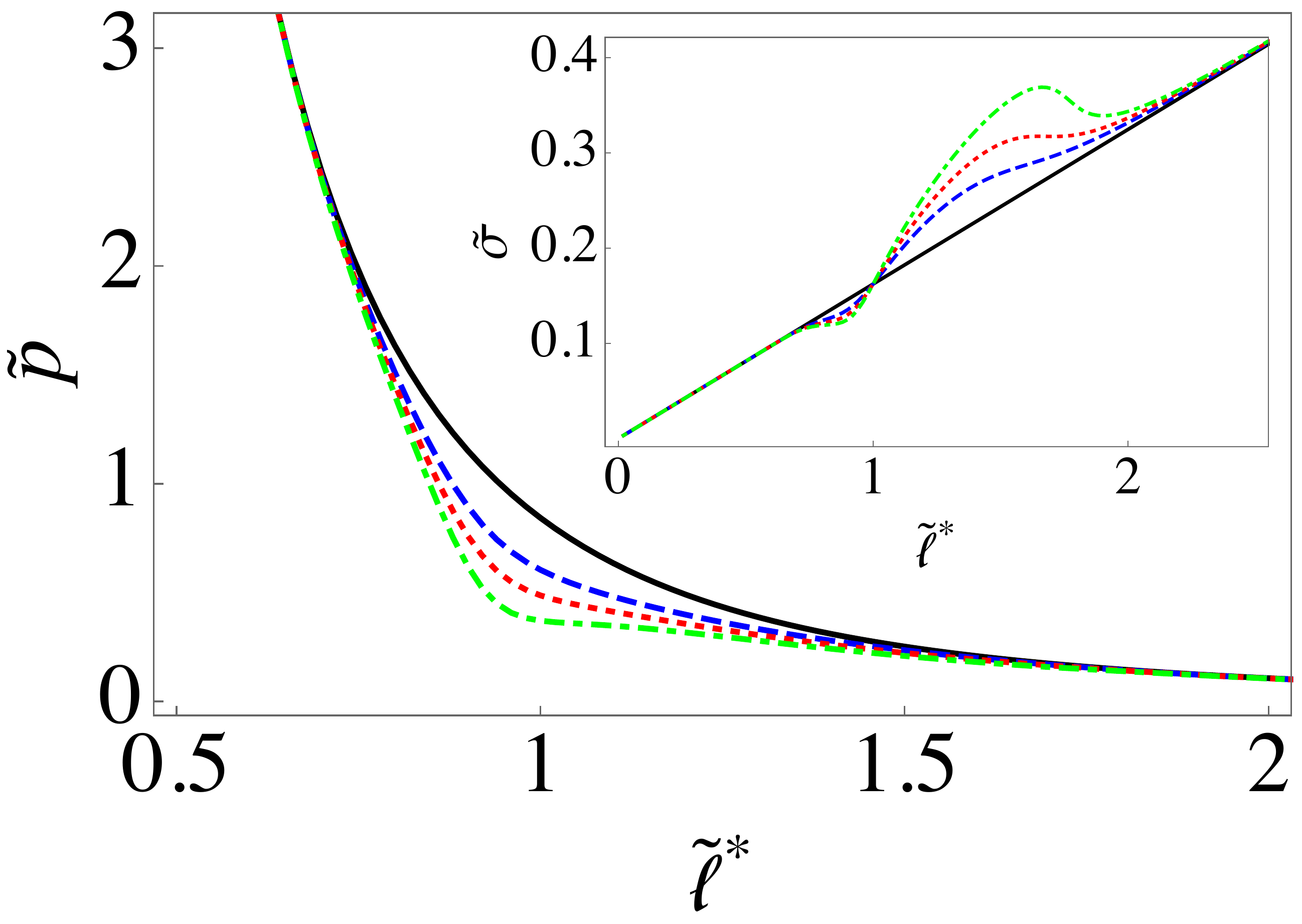}
	\caption{Attractive square well. Behavior of external osmotic pressure $\widetilde{p}$ (scaled in units of $\boltzmann T/b$) versus average separation $\tls$ for $t=0.1$ and $c=1$, and the following four binding strengths: (i)~$\widetilde{v}_0 = 0$ (black), (ii)~$\widetilde{v}_0 = -0.04$ (blue dashed), (iii)~$\widetilde{v}_0 = -0.06$ (red dotted), and (iv)~$\widetilde{v}_0 = -0.08$ (green dot-dashed). Inset: behavior of dimensionless rms fluctuation $\widetilde{\sigma} \equiv \sigma/b$ (where $b$ is the well width) versus dimensionless separation $\tls\equiv \ell_0^\ast/b$ for the same values of $t$, $c$, and the above four binding strengths.} 
\label{fig:f_attract}
\end{figure}
The variational free energy [cf. Eq.~(\ref{eq:varF})] is given by 
\ba
\label{eq:varF_square_well}
f_{{\rm var}} &\!\!=\!\!&  \frac{V_0}{2}\left( \text{erf}\left( \frac{\ell_0}{\sqrt{2}\sigma} \right) - 
 \text{erf}\left( \frac{\ell_0 - \width}{\sqrt{2}\sigma} \right) \right) 
\\
&&
+ \frac{(\boltzmann T)^2}{\bing{128}\stiff\sigma^2} + \frac{9 \boltzmann T c^2 \sigma^2 }{8 \ell_0^4} + 
 \frac{3 \boltzmann T c^2}{8 \ell_0^2} + P\ell_0. 
 \nonumber
\ea
We define a dimensionless separation distance $\widetilde{\ell}_0 \equiv \ell_0/b$, dimensionless mean square fluctuation $\widetilde{\sigma} \equiv \sigma/\width$, dimensionless well depth $\tv_0\equiv V_0 \width^2/\sqrt{2\pi} \boltzmann T$, dimensionless temperature $t \equiv \boltzmann T/\stiff$, 
%$\alpha\equiv 9c^2/4$, 
and dimensionless external osmotic pressure $\tp \equiv P\width^3/\boltzmann T$. 
In terms of these quantities, the dimensionless free energy is given by
\ba
&&\tf \equiv \beta b^2 f_{{\rm var}} 
= \frac{\sqrt{2\pi}\tv_0}{2}\left( \erf\left( \frac{\widetilde{\ell}_0}{\sqrt{2} \widetilde{\sigma}} \right) 
- \erf\left( \frac{\widetilde{\ell}_0-1}{\sqrt{2} \widetilde{\sigma}} \right)\right) 
\nonumber\\
&&\quad\quad\quad\quad+ \frac{t}{\bing{128} \widetilde{\sigma}^{2}} + \frac{9 c^2 \widetilde{\sigma}^2}{8 \widetilde{\ell}_0^{4}} + \frac{3 c^2}{8 \widetilde{\ell}_0^{2}} + \tp \, \widetilde{\ell}_0.
\label{eq:varF_square_well_dimensionless}
\ea
In the expression for the free energy above, $\sigma$ is not independent of $\ell_0$. From Eq.~(\ref{eq:implicit_relation}), we find that $\sigma$ is related to $\ell_0$ via
\ba
&&\frac{V_0}{\sqrt{2\pi}\sigma^2}\left( (\ell_0-\width)e^{-\frac{(\ell_0-\width)^2}{2\sigma^2}}
- \ell_0 e^{-\frac{\ell_0^2}{2\sigma^2}} \right)
\nonumber\\
&&\quad-
\frac{(\boltzmann T)^2}{\bing{64}\stiff \sigma^3} + \frac{9 \boltzmann T c^2 \sigma}{4 \ell_0^4} = 0     
\label{eq:sigma_square_well}
\ea 
In dimensionless form, the above equation becomes
\be
\tv_0 \big( \widetilde{\ell}_0 e^{-\frac{\widetilde{\ell}_0^2}{2 \widetilde{\sigma}^2}} -(\widetilde{\ell}_0 - 1) e^{-\frac{(\widetilde{\ell}_0 - 1)^2}{2 \widetilde{\sigma}^2}} \big) + \frac{t}{\bing{64} \widetilde{\sigma}} - \frac{9 c^2 \widetilde{\sigma}^3}{4 \widetilde{\ell}_0^4} = 0.
\label{eq:sigma_square_well_dimensionless}
\ee 
The behavior of the effective interaction energy between the membranes as a function of $\ell_0$ can be determined from Eq.~(\ref{eq:varF_square_well_dimensionless}) subject to the contraint on $\sigma$ imposed by Eq.~(\ref{eq:sigma_square_well_dimensionless}). To determine the external osmotic pressure at which the membranes are maintained at a given separation, we differentiate Eq.~(\ref{eq:varF_square_well_dimensionless}) \bing{with respect to $\ell_0$} and set it to zero (which is equivalent to making the saddle-point approximation $\langle \ell_0 \rangle_1 \approx \ell_0^\ast$). This yields
\be
\tp = \frac{3}{4(\tls)^3} + \frac{9 \widetilde{\sigma}^2}{(\tls)^5} - \frac{\tv_0}{\widetilde{\sigma}}
\Big( 
e^{-\frac{(\tls)^2}{2\widetilde{\sigma}^2}}  
- e^{-\frac{(\tls - 1)^2}{2\widetilde{\sigma}^2}} 
\Big).
\ee
\bing{For $t=0.1$ and $c=1$ the behavior of the rescaled osmotic pressure $\tp$ versus the separation $\tls$ and the behavior of the rescaled fluctuation amplitude $\tilde{\sigma}$ as a function of $\tls$ are plotted in Fig.~\ref{fig:f_attract}.}  
Note that the fluctuation amplitude is suppressed (relative to that of a purely steric system) for $\ell_0^\ast < b$, whereas it is enhanced for $\ell_0^\ast > b$. This is because for $\ell_0^\ast < b$, the membrane effectively ``sees" two repulsive (albeit one of which is finite) potential barriers and the repulsion has the effect of suppressing the amplitude of fluctuation, whereas for $\ell_0^\ast > b$, the membrane effectively ``sees" an attractive potential well which lessens the steric repulsion of the hard wall, and the fluctuation amplitude of the membrane is thus enhanced. \bing{For even larger values of $\ell_0^\ast$, the fluctuation amplitude tends towards that of a purely steric membrane (represented by the black line), as it should.} 

\bing{We note that the use of the Gaussian variational approximation is reliable for sufficiently small well depths and/or membrane separations, a criterion of reliability being that the rms fluctuation behavior has to approach that of a purely steric membrane at sufficiently large membrane separations, as the square well potential is short-ranged. For larger well depths and/or larger separations, the application of the smooth Gaussian approximation to a sharp square well results in a certain oscillatory behavior reminiscent of the Gibbs phenomenon (which can be observed in the rms fluctuation behavior at large membrane separations) and the predictions specifically for the order of the unbinding transition are not reliable. In fact the VGA-based formalism with attractive square well potential and our steric potential predicts a discontinuous unbinding transition. This discrepancy with FRG-based approaches ~\cite{lipowsky1,lipowsky2} that predict a continuous unbinding transition could arise from the relative smallness of the co-dimension (which is unity) of the membrane~\cite{mezard_parisi} and/or from the fact that the hard steric potential is being approximated by the soft(er), long(er)-range interaction of a finite $\lambda$, reflecting the size of a soft boundary region that depends on the microscopic details of the chemical make-up of the membrane~\cite{sornette}.}

\subsubsection{Morse potential}     

Let us now consider the following Morse potential, which has been used to describe the condensation of DNA molecules in multivalent salts~\cite{todd,hoang} and to model the interactions of fluid membranes~\cite{manghi}:
\begin{subequations}
\ba
&&\tw(\ell) = w_1(\ell) + w_2(\ell),
\\
\label{eq:w1}
&&w_1(\ell(\xvp)) = R \, e^{-\kappa_{{\rm D}}\ell(\xvp)},
\\
\label{eq:w2}
&&w_2(\ell(\xvp)) = - A \, e^{-\frac{1}{2}\kappa_{{\rm D}}\ell(\xvp)}.
\ea
\label{eq:w}
\end{subequations}
The Morse potential has three fitting parameters: the strength of (short-range) repulsion $R$, strength of (longer-range) attraction $A$, and an inverse length-scale $\kappa_{{\rm D}}$. Applying the variational approximation to $w_{\sigma^2}$ [with the aid of Eqs.~(\ref{eq:w_var}) and (\ref{eq:w})], we obtain
\ba
\tw_{\sigma^2} &\!\!=\!\!& 
 \int_{-\infty}^{\infty}\!\!\!\frac{d\ell}{\sqrt{2\pi\sigma^2}} 
 \big(
 R \, e^{-\kappa_{{\rm D}} \ell} 
- A \, e^{-\frac{1}{2}\kappa_{{\rm D}} \ell}
 \big) \,
 e^{-\frac{(\ell-\ell_0)^2}{2\sigma^2}}
 \nonumber\\
 &\!\!=\!\!&
 R \, e^{-\kappa_{{\rm D}} \ell_0+\frac{1}{2}\kappa_{{\rm D}}^2\sigma^2} 
- A \, e^{-\frac{1}{2}\kappa_{{\rm D}} \ell_0+\frac{1}{8}\kappa_{{\rm D}}^2\sigma^2}.
\ea
The variational free energy per unit area [defined in Eq.~(\ref{eq:varF})] becomes 
\ba
f_{{\rm var}} &\!\!=\!\!&  R \, e^{-\kappa_{{\rm D}} \ell_0+\frac{1}{2}\kappa_{{\rm D}}^2\sigma^2} 
- A \, e^{-\frac{1}{2}\kappa_{{\rm D}} \ell_0+\frac{1}{8}\kappa_{{\rm D}}^2\sigma^2} 
\\
&&
+ \frac{(\boltzmann T)^2}{\bing{128}\stiff \sigma^2} + \frac{9 \boltzmann T c^2 \sigma^2 }{8 \ell_0^4} + 
 \frac{3 \boltzmann T c^2}{8 \ell_0^2} + P\ell_0. 
\nonumber
%\label{eq:wsigmasquare}
\ea
Defining dimensionless quantities $u_1 \equiv \beta \kappa_{{\rm D}}^{-2} R$, $u_2 \equiv \beta \kappa_{{\rm D}}^{-2} A$, $\widetilde{f} \equiv \beta \kappa_{{\rm D}}^{-2} f_{{\rm var}}$, $\tl \equiv \kappa_{{\rm D}} \ell_0$, $\widetilde{\sigma} \equiv \kappa_{{\rm D}} \sigma$, $\tp \equiv \beta \kappa_{{\rm D}}^{-3} P$, and $t \equiv \boltzmann T/\stiff$, we can recast the above free energy in dimensionless form:
\ba
\widetilde{f} &=& u_1 e^{-\tl+\frac{1}{2}\widetilde{\sigma}^2} - u_2 e^{-\frac{1}{2}\tl + \frac{1}{8}\widetilde{\sigma}^2}
+\frac{3c^2}{8\tl^2}
\nonumber\\
&&+\frac{t}{\bing{128}\widetilde{\sigma}^2} + \frac{9c^2 \widetilde{\sigma}^2}{8\tl^4} + \tp \tl. 
\ea
    \begin{figure}
	\centering
		\includegraphics[width=0.46\textwidth]{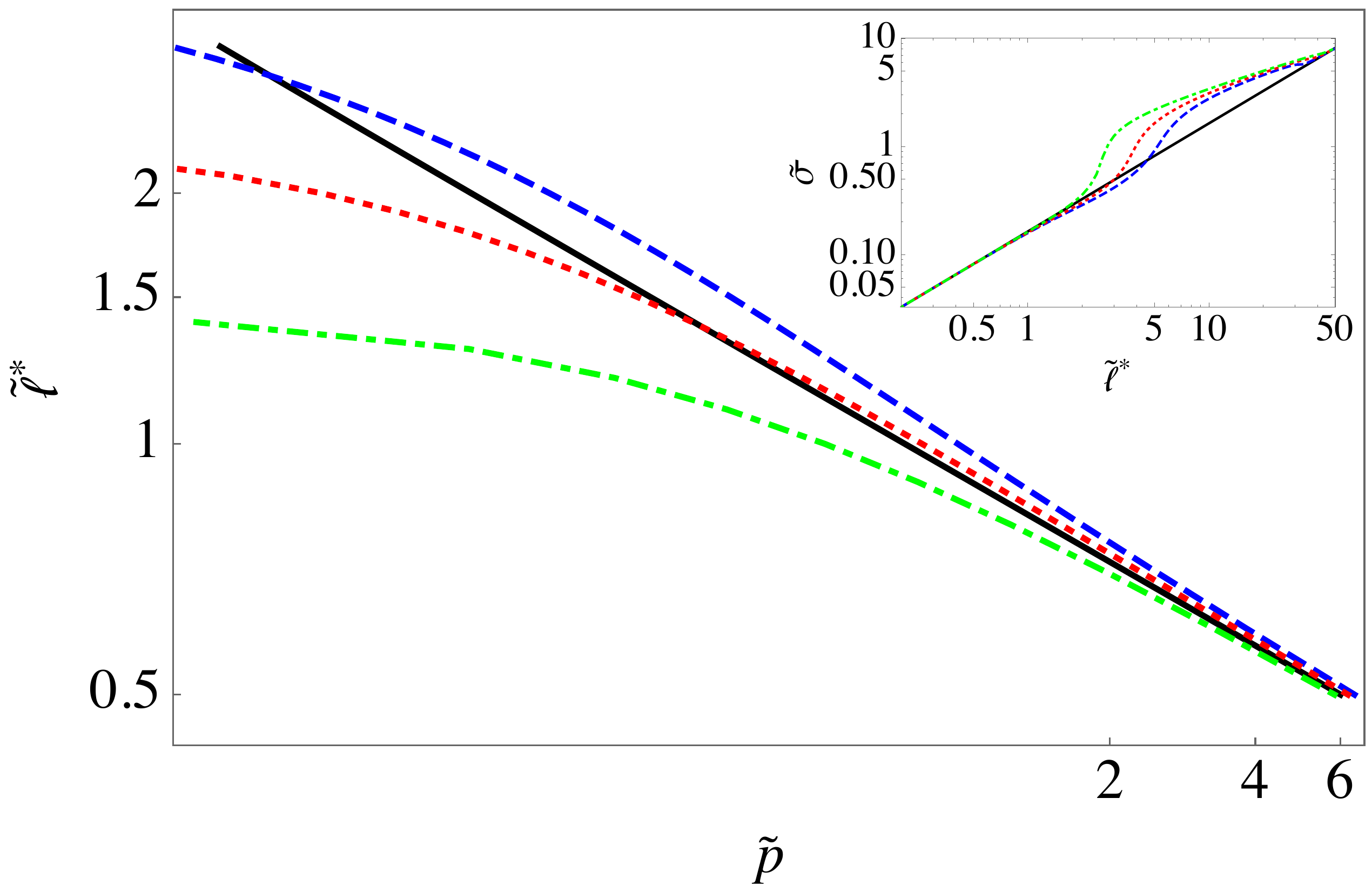}
	\caption{Morse potential. Behavior of $\tls \equiv \kappa_{{\rm D}}\ell_0^\ast$ (where $\ell_0^\ast$ is the saddle-point approximation to $\langle \ell_0 \rangle_1$) as a function of external osmotic pressure $\tp\equiv \beta P \kappa_{{\rm D}}^{-3}$ and (inset) behavior of rms fluctuation amplitude $\widetilde{\sigma} \equiv \kappa_{{\rm D}}\sigma$ as a function of  separation $\tls$ for $c=1$, $\stiff = 10\boltzmann T$, $\tls \ll \sqrt{S}$, and the following four cases: (i)~$u_1=u_2=0$ (black), (ii)~$u_1=1, u_2=0.5$ (blue, dashed), (iii)~$u_1=u_2=1$ (red, dotted), and (iv)~$u_1=1, u_2=2$ (green, dot-dashed).}
\label{fig:morse}
\end{figure}
By varying $\widetilde{f}$ with respect to $\widetilde{\sigma}$, we obtain a self-consistent relation:
\be
%&&
\frac{u_1}{2} e^{-\tl + \frac{1}{2}\widetilde{\sigma}^2} 
- \frac{u_2}{8} e^{-\frac{1}{2} \tl + \frac{1}{8}\widetilde{\sigma}^2} 
%\nonumber\\
%&&\quad
-
\frac{t}{\bing{128} \widetilde{\sigma}^4} + \frac{9 c^2}{8 \tl^4} = 0.    
\ee 
In the saddle-point approximation, the equilibrium separation $\langle \ell_0 \rangle$ is given by the solution $\tl=\tls$ to $\partial \widetilde{f}/\partial \tl = 0$:
\be
\tp = u_1 e^{-\tl+\frac{1}{2}\widetilde{\sigma}^2} - \frac{u_2}{2} e^{-\frac{1}{2}\tl + \frac{1}{8}\widetilde{\sigma}^2}
%\nonumber\\
%&&
+ \frac{9c^2 \widetilde{\sigma}^2}{2\tl^5} +\frac{3c^2}{4\tl^3}.
\ee
In Fig.~\ref{fig:morse} the behavior of $\tls$ as a function of $\tp$ and the behavior of $\widetilde{\sigma}$ as a function of $\tls$ are plotted for $c=1$, $\stiff = 10\boltzmann T$, $\tls \ll \sqrt{S}$, and the following four cases: (i)~$u_1=u_2=0$ (black), (ii)~$u_1=1, u_2=0.5$ (blue, dashed), (iii)~$u_1=u_2=1$ (red, dotted), and (iv)~$u_1=1, u_2=2$ (green, dot-dashed). Note that for cases (ii) and (iii), the rms fluctuation of the membrane is smaller than the steric-only case (i) for $\tls \lesssim 1$, as the repulsive interaction dominates in this range of separations. At separations larger than $\tls \sim 1$, the attractive $u_2$ interaction dominates over the repulsive $u_1$ interaction, and the rms fluctuation is enhanced relative to the steric-only case, becoming larger for larger attraction strengths $u_2$. At still larger separations the rms fluctuation of all four cases converge as the steric potential (which decays as $(\tls)^{-2}$) dominates over the exponentially decaying attractive tail. The rms fluctuation goes to zero at $\tls=0$ owing to the hard wall constraint. For $u_2 > 0$ the membranes are always bound at zero external osmotic pressure, and the bound state separation is smaller for larger values of $u_2$.

\subsubsection{Hydration and van der Waals forces}
%\begin{figure}
	%	\includegraphics[width=0.4\textwidth]{vdw_hyd_s_vs_ell.pdf}
	%\caption{Plot of root mean square fluctuation $\sigma$ (vertical axis, units of $\lambda$) vs mean separation $\ell_0$ (horizontal axis, units of $\lambda$).}
%\label{fig:vdw_hyd_s_vs_ell}
%\end{figure}
\begin{figure}
		\includegraphics[width=0.48\textwidth]{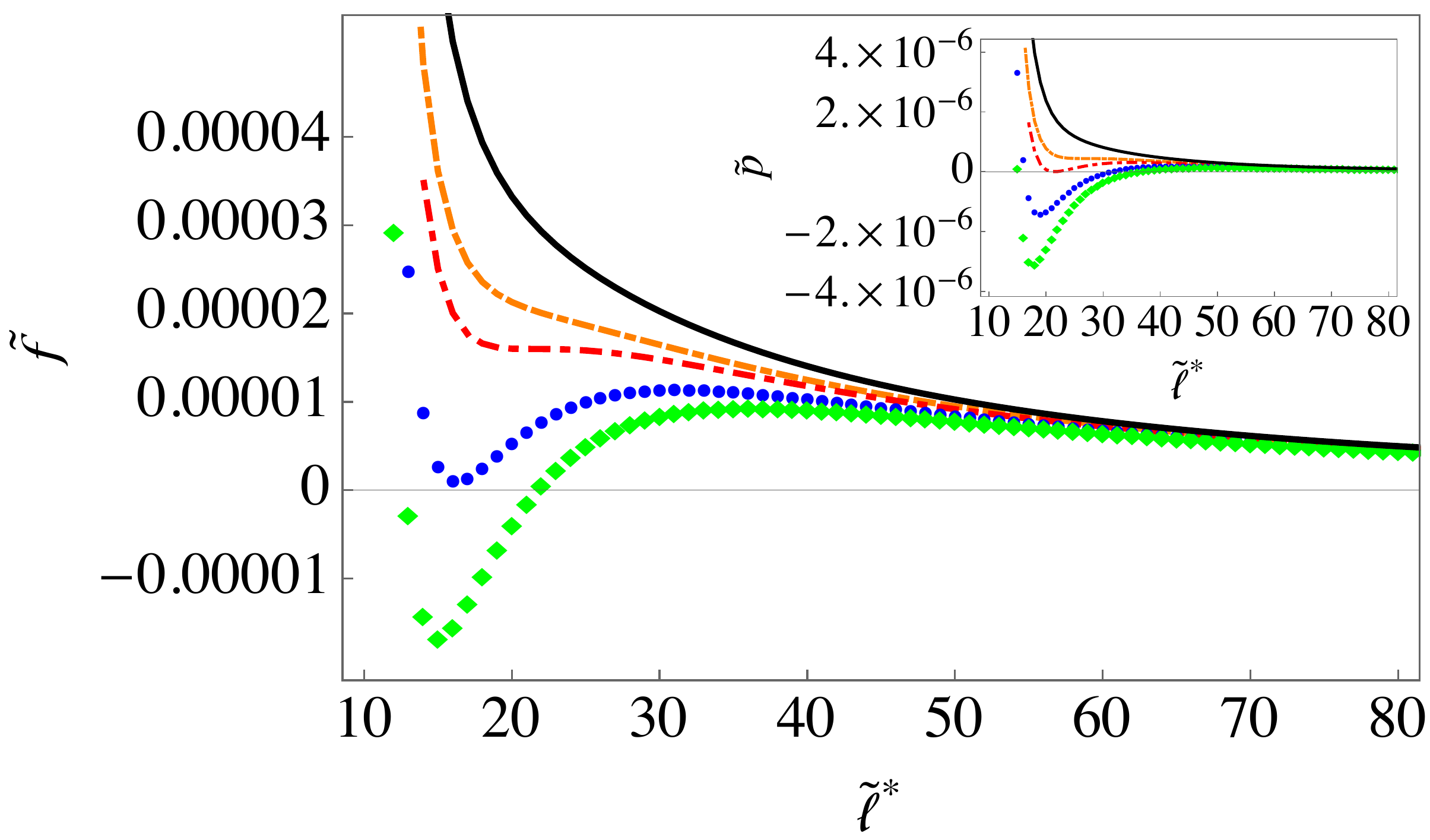}
	\caption{Hydration repulsion and van der Waals attraction. Behavior of dimensionless free energy $\widetilde{f} \equiv \beta \lambda_{{\rm H}}^2 f_{\rm{var}}$ (with $P=0$) and (inset) dimensionless external osmotic pressure $\wp$ with respect to rescaled separation $\tls \equiv \ell_0^\ast/\lambda_{{\rm H}}$ for $\widetilde{A}_{{\rm H}}=\beta A_{{\rm H}} \lambda_{{\rm H}}^2=4.83$, $\widetilde{w} = \beta W/12\pi$, $\widetilde{\delta} = \delta/\lambda_{{\rm H}} = 13.3$, $t=\boltzmann T/K = 0.0248$, $T=270 K$, $c=0.255$, and the following Hamaker strengths: (i)~$\widetilde{w}=0.04$ (black), (ii)~$\widetilde{w}=0.04902$ (orange dash-dash-dotted), (iii)~$\widetilde{w}=0.053$ (red dot-dashed), (iv)~$\widetilde{w}=0.061$ (blue disks), and (v)~$\widetilde{w}=0.068$ (green diamonds).}
\label{fig:hydration}
\end{figure}

Let us now consider the case of two uncharged membranes interacting via hydration and van der Waals forces, a generic case of interacting zwitterionic lipid membranes \cite{rand_parsegian,petrache}. For a pair of membranes of finite thickness $\delta$, the interaction potential can be expressed as~\cite{wennerstrom,petrache} 
\be
\widetilde{w}(\ell) = A_{{\rm H}}e^{-\frac{\ell}{\lambda_{{\rm H}}}} - \frac{W}{12\pi}\Big\{ \frac{1}{\ell^2} - \frac{2}{(\ell + \delta)^2} + \frac{1}{(\ell + 2\delta)^2} \Big\}, 
\ee
where the first term describes repulsion due to hydration forces and the second term describes van der Waals attraction of Hamaker strength $W$. 
For small membrane undulations, we can expand $\ell(\xvp)$ around $\ell_0$ to quadratic order in $\delta\ell(\xvp)$. Using Eqs.~(\ref{eq:w_var}) and (\ref{eq:varF}), we obtain
\ba
&&\beta f_{\rm{var}} = \frac{\boltzmann T}{\bing{128} \stiff \sigma^2} + \frac{3c^2}{8\ell_0^2} + \beta A_{{\rm H}} e^{-\frac{\ell_0}{\lambda_{{\rm H}}}+\frac{\sigma^2}{2\lambda_{{\rm H}}^2}} + \beta P \ell_0
\nonumber\\
&&\qquad+\left\{ \frac{9c^2}{8\ell_0^4} - \frac{\beta W}{4\pi} \left[ \frac{1}{\ell_0^4} - \frac{2}{(\ell_0 + \delta)^4} + \frac{1}{(\ell_0 + 2\delta)^4} \right] \right\}\sigma^2
\nonumber\\
&&\qquad- \frac{\beta W}{12\pi} \left\{ \frac{1}{\ell_0^2} - \frac{2}{(\ell_0 + \delta)^2} + \frac{1}{(\ell_0 + 2\delta)^2} \right\}
\ea
Defining rescaled quantities $t \equiv \boltzmann T/\stiff$, $\widetilde{\ell}_0 \equiv \ell_0/\lambda_{{\rm H}}$, $\widetilde{\delta} \equiv \delta/\lambda_{{\rm H}}$, $\widetilde{A}_{{\rm H}} \equiv \beta A_{{\rm H}} \lambda_{{\rm H}}^2$, $\tp \equiv \beta P \lambda_{{\rm H}}^2$, $\widetilde{w} \equiv \beta W/12\pi$ and $\widetilde{f} \equiv \beta \lambda_{{\rm H}}^2 f_{\rm{var}}$, we can re-express the above equation in dimensionless form:
\ba
&&\widetilde{f} = \frac{t}{\bing{128} \widetilde{\sigma}^2} + \frac{3c^2}{8\widetilde{\ell}_0^2} + \widetilde{A}_{{\rm H}} e^{-\widetilde{\ell}_0+\frac{1}{2}\widetilde{\sigma}^2} + \widetilde{p} \, \widetilde{\ell}_0
\nonumber\\
&&\qquad+\left\{ \frac{9c^2}{8\widetilde{\ell}_0^4} - 3\widetilde{w} \left[ \frac{1}{\widetilde{\ell}_0^4} - \frac{2}{(\widetilde{\ell}_0 + \widetilde{\delta})^4} + \frac{1}{(\widetilde{\ell}_0 + 2\widetilde{\delta})^4} \right] \right\}\widetilde{\sigma}^2
\nonumber\\
&&\qquad- \widetilde{w} \left\{ \frac{1}{\widetilde{\ell}_0^2} - \frac{2}{(\widetilde{\ell}_0 + \widetilde{\delta})^2} + \frac{1}{(\widetilde{\ell}_0 + 2\widetilde{\delta})^2} \right\}
\ea
The relation between $\sigma^2$ and $\ell_0$ is given by Eq.~(\ref{eq:implicit_relation}), which yields
\ba
&&\widetilde{A}_{{\rm H}} e^{-\widetilde{\ell}_0+\frac{1}{2}\widetilde{\sigma}^2} - \frac{t}{\bing{64} \widetilde{\sigma}^4} + \frac{9c^2}{4\widetilde{\ell}_0^4} 
\\
&&\quad- 6\widetilde{w} \left\{ \frac{1}{\widetilde{\ell}_0^4} - \frac{2}{(\widetilde{\ell}_0 + \widetilde{\delta})^4} + \frac{1}{(\widetilde{\ell}_0 + 2\widetilde{\delta})^4} \right\} = 0.     
\nonumber
\ea 
In Fig.~\ref{fig:hydration}, we study the behavior of the free energy and external osmotic pressure as functions of inter-membrane separation, for fixed hydration strength $A_{{\rm H}} = 0.2 \, {\mathrm{J \, m}}^{-2}$, hydration lengthscale $\lambda_{{\rm H}} = 0.3 \, {\mathrm{nm}}$, and bilayer thickness $\delta = 4 \,{\mathrm{nm}}$.  The critical point $\widetilde{w}_0 = \widetilde{w}_c$ at which $\partial \tp/\partial \tls = 0$ and $\partial^2 \tp/\partial (\tls)^2 = 0$ is determined numerically; we find that $\widetilde{w}_c = \bing{0.04902}$ for $t=0.0248$ and $c=0.255$. We also find the critical separation $\ell_c =\bing{26.9}\lambda_{{\rm H}}$ and the critical rms fluctuation amplitude $\sigma_c = \bing{4.76}\lambda_{{\rm H}}$. For Hamaker strengths greater than $\widetilde{w}_c$ the system exhibits phase co-existence similar to the one found in the case of the attractive square well (Sec.~\ref{sec:attractive_square_well}). The threshold Hamaker strength at which the membranes undergo a discontinuous and complete unbinding transition at zero external osmotic pressure is estimated to be $\widetilde{w}_d=\bing{0.061}$ (the blue disks in Fig.~\ref{fig:hydration}). 

Whereas FRG-based approaches~\cite{lipowsky1,lipowsky2} predicted a continuous unbinding transition for a pair of steric membranes interacting via van der Waals (vdw) and hydration forces, our VGA-based formalism with our steric potential predicts a \emph{discontinuous} unbinding transition. 
\bing{Just as we concluded for the short-ranged square-well potential in this case too the VGA-based formalism as implemented here does not reliably predict the order of the unbinding transition when compared with the FRG-based approaches~\cite{lipowsky1,lipowsky2} for the same reasons as already invoked above.}

\section{Summary and Discussion}
We have proposed a self-consistent theory for studying the interaction between a pair of mutually impenetrable and thermally undulating fluid membranes giving in the process a definitive and consistent form to the previous partially successful attempts in the same direction \cite{evans_parsegian,podgornik_parsegian}. 

We have implemented the steric constraint via the Panyukov-Rabin representation of the Heaviside function. For a pair of membranes of bending stiffness $\stiff$ and cross-sectional area $S$ interacting exclusively via steric and fluctuation forces, and separated by a mean distance $\ell_0$, we have derived a \bing{closed-form} expression for the steric potential per unit area $V_s$ [see Eq.~(\ref{eq:fs_exact})]:
\ba
V_s &=& \frac{3\boltzmann T c^2}{8\ell_0^2} 
-\frac{\boltzmann T}{\bing{8}\pi S}\ln\left\{ 1+\frac{9\boltzmann T c^2 S^2}{4\stiff \ell_0^4} \right\} 
\nonumber\\
&&+ \frac{\boltzmann T}{\bing{4}\pi} \sqrt{\frac{9\boltzmann T c^2}{4\stiff \ell_0^4}}
\bigg\{
\tan^{-1}\bigg( 1-\frac{\sqrt{2}(4\stiff)^{1/4}\ell_0}{(9\boltzmann T c^2)^{1/4}\sqrt{S}} \bigg)
\nonumber\\
&&\qquad+
\tan^{-1}\bigg( 1+\frac{\sqrt{2}(4\stiff)^{1/4}\ell_0}{(9\boltzmann T c^2)^{1/4}\sqrt{S}} \bigg)
\bigg\}.
\ea
This has two contributions: one that is induced by zero mode fluctuations of the membranes and one that is induced by thermal bending fluctuations. At small separations $\ell_0 \ll \sqrt{S}$, the bending fluctuation-dependent part scales as $\ell_0^{-2}$, and crosses over to $\ell_0^{-4}$ scaling for large separations $\ell_0 \gg \sqrt{S}$. On the other hand, the zero mode-dependent part of the steric potential always scales as $\ell_0^{-2}$. Concomitantly we also derived a \bing{closed-form} formula [see Eq.~(\ref{eq:sigma_exact})] for the rms undulation amplitude $\sigma$, viz.,
\be
\sigma^2 = \frac{\ell_0^2}{\bing{12}c}\sqrt{\frac{\boltzmann T}{\stiff}}\left( 1 - \frac{2}{\pi}\tan^{-1}\left( \frac{2\ell_0^2}{3cS}\sqrt{\frac{\stiff}{\boltzmann T}} \right) \right),
\ee
which has the following asymptotic behavior: for $\ell_0 \ll \sqrt{S}$, $\sigma$ scales linearly with $\ell_0$, whereas for $\ell_0 \gg \sqrt{S}$, $\sigma$ saturates at the order of $\boltzmann T S/\stiff$. The rms fluctuation amplitude becomes small at low temperatures $T$ and/or large bending stiffnesses $\stiff$. We believe that our result refines and substantiates the Ansatz $\sigma^2 = \hat{\mu} \ell_0^2$ first postulated (on the basis of heuristic arguments) in Ref.~\cite{helfrich}. 
 
To investigate fluid membrane systems that experience interactions of non-Gaussian form, we have adapted the Feynman-Kleinert version of the variational Gaussian approximation (VGA) to the case of fluid membranes subject to our effective steric potential $V_s$. We have applied this VGA approach to four different types of potential: (i)~the Moreira-Netz potential for a pair of strongly charged membranes with an intervening solution of multivalent counterions, (ii)~an attractive square well, (iii)~the Morse potential, and (iv)~a combination of hydration and van der Waals interactions. 

In the first case we make a prediction that, everything else being the same, the multivalent counterion-mediated interaction measured between hard and soft surfaces should display a substantial difference in the equilibrium spacing, with soft surfaces displaying larger equilibrium spacing. This difference should increase with the valency of the counterions and could easily reach a factor of 2 for high valency counterions and should thus be eminently measurable in the planned osmotic-stress experiments with lipid membranes in the presence of multivalent salts \cite{Pabst}. We furthermore note here that our results for multivalent salts imply also a pronounced effect on the estimated values of the bending rigidities that could be extracted from the linewidth of the X-ray scattering intensity in this type of experiments, irrespective of whether they directly renormalize the bending rigidity of lipid membranes or not \cite{Mathias, Siggia, Kumaran}. 

Within the same VGA approach we also analyzed in detail the three other cases of the coupling between long-range interactions and conformational fluctuations, showing the versatility and usefulness of our approach in the context of widely differing types of long-range interactions in the regime of intermediate separations, which are also the typical length scales of biologically relevant systems. 

As future projects in the same general direction, our self-consistent theory can be employed to study multi-component membranes, membranes with non-zero surface tension, and tethered and polymerized membranes.  

\section{Acknowledgment}
The authors thank A. C. Maggs, G. Pabst, D. J. G. Crow, P. Ziherl, A. Siber, B. Kollmitzer and S. Gupta for rewarding discussions and acknowledge the financial support of the Agency for research and development of Slovenia (ARRS) under the bilateral SLO-A grant  N1-0019.

\end{document}